\documentclass[prd,aps,amsfonts,eqsecnum,superscriptaddress,nofootinbib,longbibliography,notitlepage]{revtex4-1}

\usepackage{graphicx}
\usepackage{xcolor}
\usepackage{rotating}
\usepackage{amsmath,amssymb,graphics,amsthm,isomath}
\usepackage{bbm}
\usepackage{bm}
\usepackage{array}
\newcolumntype{P}[1]{>{\centering\arraybackslash}p{#1}}
\usepackage{multirow}

\usepackage[colorlinks=true, urlcolor=violet, linkcolor=blue, citecolor=red, hyperindex=true, linktocpage=true]{hyperref}
\usepackage[capitalise,compress]{cleveref}

\allowdisplaybreaks

\numberwithin{thm}{section}

\makeatletter
\renewcommand{\p@subsection}{}
\renewcommand{\p@subsubsection}{}
\makeatother

\usepackage{xcolor}
\usepackage{mathtools}

\newcommand\bea{\begin{eqnarray}}
\newcommand\eea{\end{eqnarray}}
\newcommand\be{\begin{equation}}
\newcommand\ee{\end{equation}}
\newcommand\bes{\begin{subequations}}
\newcommand\ees{\end{subequations}}
\newcommand\bed{\begin{displaymath}}
\newcommand\eed{\end{displaymath}}
\newcommand\beal{\begin{aligned}}
\newcommand\eeal{\end{aligned}}
\newcommand\bew{\begin{widetext}}
\newcommand\eew{\end{widetext}}
\newcommand\beit{\begin{itemize}}
\newcommand\eeit{\end{itemize}}
\def\bea{\begin{array}}
\def\eea{\end{array}}
\newcommand\been{\begin{enumerate}}
\newcommand\eeen{\end{enumerate}}

\newcommand{\eqnref}[1]{\eqref{#1}}

\newcommand{\tabref}[1]{Table\;\ref{#1}}
\newcommand{\secref}[1]{Section\;\ref{#1}}
\newcommand{\appref}[1]{Appendix\;\ref{#1}}

\newcommand{\ud}{\mathrm{d}}
\def\i{\text{i}}

\newcommand\mT{\mathcal{T}}

\newcommand\bbZ{\mathbb{Z}}

\newcommand\mO{\mathcal{O}}
\newcommand\mP{\mathcal{P}}

\newcommand\mL{\mathcal{L}}
\newcommand{\mfb}{\mathfrak{b}}
\newcommand{\mfc}{\mathfrak{c}}
\newcommand{\mfw}{\mathfrak{w}}
\newcommand\mrmD{\mathrm{D}}

\newcommand{\vect}[1]{\boldsymbol{#1}}
\newcommand{\im}{\mathrm{Im}~}

\newcommand{\phan}{\phantom{i}}

\newcommand{\p}{\partial}

\usepackage{dsfont}

\begin{document}

\title{Hydrodynamic effective field theories with discrete rotational symmetry}

\author{Xiaoyang Huang}
\email{xiaoyang.huang@colorado.edu}
\affiliation{Department of Physics and Center for Theory of Quantum Matter, University of Colorado, Boulder CO 80309, USA}

\author{Andrew Lucas}
\email{andrew.j.lucas@colorado.edu}
\affiliation{Department of Physics and Center for Theory of Quantum Matter, University of Colorado, Boulder CO 80309, USA}

\date{\today}

\begin{abstract}
    We develop a hydrodynamic effective field theory on the Schwinger-Keldysh contour for fluids with charge, energy, and momentum conservation, but only discrete rotational symmetry.   The consequences of anisotropy on thermodynamics and first-order dissipative hydrodynamics are detailed in some simple examples in two spatial dimensions, but our construction extends to any spatial dimension and any rotation group (discrete or continuous). We find many possible terms in the equations of motion which are compatible with the existence of an entropy current, but not with the ability to couple the fluid to background gauge fields and vielbein.
\end{abstract}

\maketitle

\tableofcontents

\section{Introduction}

Recent years have seen a flurry of activity to develop a dissipative effective field theory for hydrodynamics \cite{SonPRD, Eightfold1, Eightfold2, Crossley2017, Glorioso2017, Liulec}.  While the actual field theory itself was well-known for a very long time (e.g. the stochastic Navier-Stokes equations), what remained mysterious were the underlying symmetries that might lead to the explicit construction of a Lagrangian, as one does in textbook quantum effective field theories (for non-dissipative dynamics).  

Thus far, many of the papers written on this subject have sought to understand \emph{existing} hydrodynamic universality classes, including the Navier-Stokes equations, in a new field theoretic language.  More recently, however, some authors have begun to use this effective field theoretic approach to predict new universality classes of hydrodynamics, such as ``fracton fluids" \cite{GromovPRR, breakdown,grosvenor2021hydrodynamics} that arise in constrained quantum dynamics.

The purpose of this paper is to use effective field theory methods to learn about ``regular" fluids (with charge, energy and momentum conservation) with only discrete rotational symmetry groups.  This is particularly relevant for applications to electron liquids in high-purity materials \cite{gurzhi,Guo_2017,Levitov_2016,Torre_2015,alekseev,Andreev_2011,Forcella_2014,Tomadin_2014,lucas16,ephh}, which have been realized experimentally in many materials \cite{vool2020imaging,Bandurin_2016,Crossno_2016,ghahari,Krishna_Kumar_2017,Gallagher158,Berdyugin_2019,sulpizio,jenkins2020imaging,Ku_2020,fu2018thermoelectric,Moll_2016,Gusev_2018}; see \cite{Lucas_2018} for a review. Since most metals are not even close to isotropic, it is important to understand the consequences for discrete rotational symmetry on hydrodynamics.  Some literature \cite{link,Cook2019, Cook2021, Narang, BradlynPRX}  has already attempted to describe the hydrodynamics of such anisotropic fluids, albeit usually by simply positing the allowed tensor structures that could arise in (e.g.) viscosity.  When done, authors have used kinetic theory \cite{link,Cook2019} or AdS/CFT \cite{Rebhan:2011vd,Jain:2015txa,Blake:2016wvh} to derive the equations for an anisotropic fluid from a more microscopic perspective.  

As we will see, there are a few peculiarities which are somewhat surprising from a microscopic perspective, and which it is desirable to have a more universal understanding of.  For example, we will see that in fluids with triangular point group \cite{triangle}, there are certain terms which \emph{seem} to be allowed in the conventional Landau paradigm (an entropy current can be constructed): in particular within linear response, it seems possible to construct a spatial stress tensor $c\cdot  \lambda_{ijk}v_k \subset T_{ij} $, with $v_k$ fluid velocity and $\lambda$ an invariant tensor under the discrete point group.  However, kinetic theory calculations reveal that $c=0$ \cite{triangle}.  In this paper, we will explain why $c=0$ in this model based on very general arguments which are most natural within the effective theory approach. In other point groups as well, we will show that certain anisotropic corrections, naively allowed by symmetry or Landau phenomenology, can be forbidden.

We will follow rather closely the formalism introduced in \cite{Crossley2017,Glorioso2017} as we develop our effective field theory of hydrodynamics.  The main difference between our work and earlier work on the subject is that for a discrete point group, there are no continuous generators of rotational symmetry whatsoever.  The key consequence of this is that, just as when one studies a non-relativistic fluid it is more appropriate to couple the fluid to an Aristotelian background \cite{Boer1, Armas2021} rather than a conventional Lorentzian spacetime manifold, here we will find it desirable to ``generalize" the Aristotelian background to an even more generic family of geometries which does not demand any accidental symmetry.  The natural conclusion is that one should couple the fluid with only discrete rotational symmetry directly to the vielbein.  The vielbein indices will encode all information about the discrete (or continuous) rotational symmetries imposed on the model.  To understand this conclusion, notice that when coupling to a relativistic metric $g_{\mu\nu}$, the stress tensor $T^{\mu\nu} \sim \delta S/\delta g_{\mu\nu}$ must manifestly be symmetric.  This means physically that energy current is the same as momentum density (in proper units).  The non-relativistic fluid can only avoid this Lorentz covariance by coupling not to $g_{\mu\nu}$ but to a ``spatial metric" $h_{\mu\nu}$ and a timelike vector $\tau_\mu$ obeying suitable constraints, similar to Newton-Cartan geometry \cite{SonNC1, SonNC2, Jain2020}.  In an anisotropic fluid, there are in general \emph{no symmetry requirements} on the stress tensor.  So the only object we can couple to is a set of $d+1$ linearly independent spacetime vectors, i.e. the vielbein.


In \secref{sec:eft}, we review the geometry of the Aristotelian background, and then generalize the effective field theory to the vielbeins, focusing on the classical limit of most relevance for hydrodynamics.  In \secref{sec:ideal}, we consider thermodynamics and the ``ideal fluid" limit, and explain why certain terms can be forbidden despite their naive compatibility with Landau's formulation of hydrodynamics based on entropy currents.  In \secref{sec:dissipative}, we discuss first order dissipative hydrodynamics and describe both fluids with discrete and continuous rotational symmetries based on our formalism. We discuss further the parity-violating hydrodynamics with applications to the Hall effect in \secref{sec:parity}. Finally, \secref{sec:conclusion} contains concluding remarks.

\section{Overview of effective field theory}\label{sec:eft}
In this section we will overview the effective field theory framework along with the symmetries we impose.

\subsection{Fields in the effective action}

Consider a generating function in the Schwinger-Keldysh (S-K) formalism \cite{Kamenev_book, SKopen}\footnote{For our purpose, we only consider the closed time path \cite{Liulec}.} for correlators of a conserved U(1) current $J^\mu$ and the energy and momentum currents $T^\mu_{\alpha}$.  Here and below, Greek $\mu\nu\cdots$ indices will represent \emph{coordinate spacetime indices}, while $ij\cdots$ represent only spatial components.   $\alpha,\beta,\ldots$ represent spacetime vielbein indices, while $b,c,\ldots$ represent spatial vielbein indices only (we will reserve $a$ for another purpose!). The summation convention is used for all four types of indices. We emphasize that for us, the energy and momentum currents are most naturally thought of as ordinary vectors, with an additional \emph{vielbein} index associated to the actual quantity being conserved.  Our goal is to calculate the generating function of real-time correlation functions of $J^\mu$ and $T^\mu_\alpha$.  This is done in the standard way by constructing the generating functional \begin{equation}\label{eq:W}
    \mathrm{e}^{W[e_{s\mu}^\alpha, A_{s\mu}]} = \mathrm{tr}\left(\rho_0 U^\dagger(e_{2\mu}^\alpha, A_{2\mu})U(e_{1\mu}^\alpha, J_{1\mu})\right),
\end{equation}
where the unitary operators $U$ are defined as \begin{equation}
    U(e_\mu^\alpha,A_\mu) = \exp\left[\mathrm{i}\int \mathrm{d}t \mathrm{d}^dx \; \left(e_\mu^\alpha(x)T^\mu_\alpha(x) + A_\mu(x)J^\mu(x)\right)\right].
\end{equation}
Here the $s$ index runs over indices 1 and 2, and denotes the half of the S-K contour on which the field is defined.  The global U(1) symmetry of the field theory implies that $W$ must be gauge invariant (in the absence of anomalies) with respect to the background gauge field $A_\mu$.


$e_\mu^{\alpha}$ is the vielbein: it will play the role of the spacetime metric in our calculation. We must use vielbein rather than a metric for two reasons. Firstly, the vielbeins can somewhat intuitively be regarded as  ``background gauge fields''\footnote{The momentum mimics the time-reversal-odd ``charges''. However, it does not diffuse but has linear dispersion relation due, in part, to the non-linear diffeomorphism symmetry we will review below. In this sense, the ``background gauge field" analogy is a bit imprecise.}, so, similar to the $\mathrm{U}(1)$ symmetry, the spacetime diffeomorphism is realized as a kind of ``gauge invariance'' of the vielbeins. Second, as we will see frequently below, in the presence of a small symmetry group, the vielbeins are more natural and fundamental ingredients to describe the spacetime. Indeed, as we explained in the introduction, a conventional metric is too strongly constrained to capture the asymmetry of the stress tensor which is inevitable in an anisotropic theory. The vielbein indices $\alpha$ are in a representation of any discrete/continuous rotational symmetries which remain in the problem.  

The vielbein satisfy the orthogonality and completeness relations
\be\label{eq:vielbeinid}
e^\mu_\alpha e^\beta_\mu = \delta^\beta_\alpha, \quad e^\mu_\alpha e^\alpha_\nu = \delta^\mu_\nu.
\ee
We denote the determinant of the vielbeins as \begin{equation}
    e = \det(e^\mu_{\alpha}).
\end{equation} Moreover, as we do not assume the existence of an absolute time \cite{Bradlyn2015} and allow for spatial dislocations \cite{Hughes2011, Hughes2015,Hughes2013}, we introduce the torsion field 
\begin{equation}
    G^{\alpha}_{\mu\nu} \equiv \p_\mu e^{\alpha}_\nu-\p_\nu e^{\alpha}_\mu + {\omega^{\alpha}}_{\beta \mu}e^\beta_\nu - {\omega^{\alpha}}_{\beta \nu}e^\beta_\mu,
\end{equation}
where ${\omega^{\alpha}}_{\beta \mu}$ is the spin connection which makes the derivatives covariant under discrete rotational symmetry.  
In the present case, only the internal space-like components ${\omega^{b}}_{c \mu}$ are nonzero.

\subsection{Fluid symmetries}

To proceed, we represent the generating function \eqnref{eq:W} in terms of a path integral.  In principle this path integral can be done over all microscopic fields, but we wish to integrate out all of the non-hydrodynamic modes.  Following \cite{Crossley2017}, the modes which we will keep are the Stuckelberg fields $X^\mu$, which we will relate to energy and momentum, and $\phi$, which we will relate to charge:
\begin{equation}\label{eq:W2I}
    \mathrm{e}^{W[e_{1,\mu}^{\alpha},A_{1,\mu};e_{2,\mu}^{\alpha},A_{2,\mu}]} =\int \mathrm{D} X_1 \mathrm{D} X_2 \mathrm{D} \phi_1 \mathrm{D} \phi_2   ~ \mathrm{e}^{\i I_{\mathrm{EFT}}[e^{\alpha}_{1,A},B_{1,A};e^{\alpha}_{2,A},B_{2,A}]}.
\end{equation}
To incorporate diffeomorphism invariance and to promote the coordinate fields $X^\mu$ to be dynamical, it is helpful to introduce another \emph{fluid spacetime} parametrized by $\sigma^A$, as we will illustrate in the following. So, in the above equation, the $A,B$ indices denote space and time ($A=t$) in the fluid spacetime, while $I,J,\ldots$ denote spatial indices alone.  The fields in $I_{\mathrm{EFT}}$ are defined as ($ s=1,2$)
\be\label{eq:fluidv}
 e^{\alpha}_{s,A}(\sigma) = \frac{\p X_s^\mu(\sigma)}{\p \sigma^A}  e^{\alpha}_{s,\mu}(\sigma),\quad B_{s,A}(\sigma) = \frac{\p X_s^\mu(\sigma)}{\p \sigma^A} A_{s,\mu}(\sigma)+ \frac{\p \phi_s(\sigma)}{\p \sigma^A},
\ee
and one can check that they are invariant under the spacetime diffeomorphism and $\mathrm{U}(1)$ gauge symmetries:
\begin{subequations}
\begin{align}
    e^{\prime\alpha}_{s,\mu}(X_s^\prime) &= \frac{\p X_s^\nu}{\p X_s^{\prime \mu}} e^{\alpha}_{s,\nu}(X_s), \quad A^{\prime}_{s,\mu}(X_s^\prime) = \frac{\p X_s^\nu}{\p X_s^{\prime \mu}} A_{s,\nu}(X_s),\quad X_s^{\prime \mu}(\sigma) = f_s^\mu(X_s(\sigma)),\\
    A^\prime_{s,\mu}(X_s) &= A_{s,\mu}(X_s) - \p_\mu \lambda_s(X_s), \quad \phi^\prime_s(\sigma) = \phi_s(\sigma)+\lambda_s(X_s(\sigma)), 
\end{align}
\end{subequations}
for arbitrary functions $f_s^\mu$ and $\lambda_s$. To describe the hydrodynamics of the charged fluids, additional symmetries need to be imposed to distinguish from other phases of matter such as solids and superfluids\footnote{However, these symmetries lack a solid derivation (see attempts based on holography in \cite{BoerHolo, GloriosoHolo}). }. First, there is a ``diagonal shift symmetry'' for each fluid element:
\begin{equation}\label{eq:shiftsymmetry}
    \phi_r \to \phi_r + \lambda(\sigma^I),\quad \phi_a\to \phi_a.
\end{equation}
This is the freedom to make an independent phase change in a fluid. Similarly, the fluid spacetime has a reparametrization symmetry both in space and time:
\begin{subequations}
\begin{equation}\label{eq:shiftspace}
    \sigma^I \to \sigma^{\prime I}(\sigma^I),\quad \sigma^t  \to \sigma^t,
\end{equation}
\begin{equation}\label{eq:shifttime}
    \sigma^t \to \sigma^{\prime t}=\sigma^t+f(\sigma^I),\quad \sigma^I \to \sigma^I.
\end{equation}
\end{subequations}
The freedom to relabel each fluid element and set their own clocks distinguishes a fluid phase from a solid phase. Note that in \eqnref{eq:shifttime} we have fixed part of the gauge freedom relative to \cite{Crossley2017} by defining the local \emph{proper} temperature to be
\begin{equation}
    T(\sigma) =2 \frac{T_0}{e^0_{1,t}+e^0_{2,t}}.
\end{equation}
This is the standard way of defining local proper temperature in a curved spacetime; the temperature is induced by the temporal Killing vector $\p X^\mu/\p \sigma^t$.  For the purposes of this paper, this gauge fixing will be useful.

On each contour $s$, it is convenient to decompose the gauge invariant variables $e^{\alpha}_{s,A}$ as 
\bes
\begin{equation}
\frac{\p X^\mu}{\p \sigma^t}\equiv bu^\mu, \quad b= \frac{\p X^\mu}{\p \sigma^t} e^{0}_\mu,
\end{equation}
\begin{equation}
\frac{\p X^\mu}{\p \sigma^I}\equiv b u^\mu v_I+\lambda_I^\mu, \quad v_I = \frac{1}{b}\frac{\p X^\mu}{\p \sigma^I} e^{0}_\mu,\quad \lambda_I^\mu=\frac{\p X^\mu}{\p \sigma^I} - v_I \frac{\p X^\mu}{\p \sigma^t},
\end{equation}
\ees
such that
\be
u^\mu  e^{0}_\mu=1,\qquad  e^{0}_\mu \lambda_I^\mu=0.
\ee
$u^\mu$ plays the role of a fluid velocity vector, $b$ will eventually relate to temperature, and $v_I$ and $\lambda^\mu_I$ denote the parts of $\partial_I X^\mu$ oriented along $e^0_\mu$ or not.  We then define (recall $b,c$ indices run only over spatial vielbein!) 
\begin{equation}
    u^{b}= u^\mu e^{b}_\mu, \quad a_{I}^{b} = \lambda_I^\mu e^{b}_\mu,
\end{equation}
and we define $a^I_{b} = \lambda^I_\mu e^\mu_{b}$ as the inverse matrix of $a^{b}_{I}$, where $\lambda^I_\mu$ is the inverse of $\lambda_I^\mu$ satisfying
\begin{equation}
    \lambda_\mu^I \lambda^\mu_J = \delta^I_J,\quad \lambda^I_\mu \lambda^\nu_I = \delta^\nu_\mu - e^\nu_{0}e_\mu^{0}.
\end{equation}
We can similarly define the fluid spatial metric \begin{equation}
    a_{IJ}\equiv a_I^b a_J^b= \lambda_I^\mu \lambda_J^\nu h_{\mu\nu}, \;\;\;\; h_{\mu\nu}=e^b_\mu e^b_\nu
\end{equation} and its inverse $a^{IJ}$, such that we can raise or lower the indices to get $\lambda_\mu^I\equiv h_{\mu\nu}a^{IJ}\lambda_J^\nu$.  Here $h_{\mu\nu}$ is the spatial part of the metric.  In the remaining part of the paper, upper (lower) fluid spatial indices are understood as being raised (lowered) by $a^{IJ}$.
Similarly, for $B_{s,A}$, we will decompose it as
\begin{equation}
    \mu = u^\mu A_{\mu}+D_t \phi, \quad  \mfb_{I}= \lambda_I^\mu A_{\mu}+D_I \phi,
\end{equation}
where $D_t = b^{-1}\p_t$ and $D_I = \p_I-v_I\p_t$. 

The degrees of freedom $b, v_I, u^\mu, \lambda_I^\mu$ are not all independent.  In particular, from
\be
\frac{\p X^\nu}{\p\sigma^I}\p_\nu \frac{\p X^\mu}{\p\sigma^t} = \frac{\p X^\nu}{\p\sigma^t}\p_\nu \frac{\p X^\mu}{\p\sigma^I},
\ee
we have 
\bes
\begin{align}
u^\mu \partial_\mu v_I &= \frac{1}{b^2} \lambda_I^\nu \p_\nu b+\frac{1}{b}u^\mu\lambda_I^\nu  G^{0}_{\mu\nu}, \\
u^\mu \partial_\mu \lambda_I^\mu &= \lambda_I^\nu \p_\nu u^\mu-u^\mu u^\nu \lambda_I^\rho G^{0}_{\nu\rho},
\end{align}
\ees

To summarize, the gauge invariant variables are
\begin{equation}
    \Phi_s = \{ b_s, \quad v_{s,I},\quad u^{b}_s,\quad  a_{s,I}^{b},\quad   \mu_s,\quad  \mfb_{s,I} \}.
\end{equation}

In order to make these variables covariant under the fluid shift symmetries \eqnref{eq:shiftspace} and \eqnref{eq:shifttime}, we first introduce the $r$/$a$ variables.  In a nutshell the $r$-variables will correspond to hydrodynamic degrees of freedom while $a$-variables correspond to fluctuations and noise.  We will always write \begin{equation}
    \Phi_r = \frac{\Phi_1+\Phi_2}{2},
\end{equation}
but it will be convenient to define the $a$-variables in a more complicated way:
\begin{align}
    b_a &= \log (b_2^{-1}b_1),\quad v_{a,I} = v_{1,I}-v_{2,I},\quad u^{b}_a = u^{b}_1-u^{b}_2,\quad \mu_a = \mu_1-\mu_2,\quad \mfb_{a,I} = \mfb_{1,I}-\mfb_{2,I},\nonumber \\
    \chi_a &= \log \det \left( a^I_{2,b}a^{b}_{1,J}\right),\quad \Xi^I_{a,J} =\log \left(\frac{ a^I_{2,b}a^{b}_{1,J}}{\det \left( a^I_{2,b}a^{b}_{1,J}\right)}\right).
\end{align}
Note that we have separated out $a_{IJ}$'s $a$-field into the traceful ($\chi_a$) and traceless $(\Xi^I_{a,J}$) components. Now, we introduce two covariant derivatives $D_t$ and $D_I$ on the fluid spacetime. For a general covariant scalar $\xi$ they are \footnote{Note that $b_r$, $V_{r,I}$ and $\mfb_{r,I}$ are not covariant objects, thus one should take \eqnref{eq:covfield} as a definition of their first derivatives.}
\begin{equation}
    D_t \xi = \frac{1}{b_r} \p_t \xi,\quad D_I \xi = \p_I \xi - v_{r,I}\p_t \xi.
\end{equation}
For the vector fields we have instead \begin{equation}\label{eq:covfield}
    V_{a,I} = b_r v_{a,I},\quad V_{r,I} = b_r v_{r,I},\quad  D_I b_r \equiv \frac{1}{b_r}(\p_I b_r -\p_t V_{r,I} ) ,\quad D_t \mfb_{r,I} \equiv \frac{1}{b_r}\p_t \mfb_{r,I}.
\end{equation}


From \eqnref{eq:W} and \eqnref{eq:W2I}, unitarity and stability require
\begin{subequations}\label{eq:Isym}
\begin{equation}
    I^*_{\mathrm{EFT}}[\Lambda_a,\Lambda_r] = -I_{\mathrm{EFT}}[-\Lambda_a,\Lambda_r],
\end{equation}
\begin{equation}\label{eq:unitary}
    \im I_{\mathrm{EFT}} \geq 0,
\end{equation}
\begin{equation}
     I_{\mathrm{EFT}}[\Lambda_a=0,\Lambda_r] = 0,
\end{equation}
\end{subequations}
where $\Lambda_{r,a}$ denote collectively the $r$-$a$ variables of both external and dynamical fields. The first equation tells that every even power of $a$-fields should be purely imaginary, and the second one says their coefficients should be non-negative. The last equation means every term should at least include one $a$-field. 

Moreover, by taking the density matrix to be the thermal ensemble $\rho_0=\mathrm{e}^{-\beta_0 H}$, the Kubo-Martin-Schwinger (KMS) condition tells that $I_{\mathrm{EFT}}$ is related to its time-reversal partner with every field getting an imaginary shift along the temporal direction. By further applying an anti-unitary symmetry $\Theta$ that is preserved by the Hamiltonian (time-reversal, possibly in combination with a spatial operation)  we obtain the symmetry
\begin{equation}\label{eq:kmscondition}
    I_{\mathrm{EFT}}[\Lambda_1,\Lambda_2] = I_{\mathrm{EFT}}[\tilde{\Lambda}_1,\tilde{\Lambda}_2],
\end{equation}
where
\begin{equation}\label{eq:tildeoperator}
    \tilde{\Lambda}_1 = \Theta \Lambda_1(t-\i \theta,\vect{x}),\quad \tilde{\Lambda}_2 = \Theta \Lambda_2(t+\i (\beta_0-\theta),\vect{x}).
\end{equation}
We will focus on the consequences of this symmetry in the limit where we can Taylor expand in $\beta_0-\theta$, though defer a detailed discussion of this classical limit to the end of this section.

Now the action in the fluid spacetime is ready to be written down.  Its schematic structure is 
\be\label{eq:action}
I_{\mathrm{EFT}} = \int \ud^{d+1}\sigma \; a b ~\mL[\Phi_r,\Phi_a],
\ee
where we defined $a = \det a^{b}_J$.  Note the Jacobian
\be
\Lambda = \det \frac{\p X}{\p \sigma} = \frac{ab}{e}.
\ee
The $\mL$ can be further expanded in the number of $a$-fields and (covariant) derivatives:
\be
\mL = \mL^{(1,0)}+\mL^{(1,1)}+\ldots+\mL^{(2,0)}+\ldots,
\ee
where $\mL^{(m,n)}$ contains $m$ factors of $a$-fields and $n$ derivatives. In this paper, we will restrict to the order $m+n\leq 2$.\footnote{Writing down higher order terms is sophisticated and does not appear particularly illuminating to us, although there are some point groups where more interesting structure will only arise at higher orders (such as $\mathrm{D}_5$).} Then, the (off-shell) stress tensor and current can be derived through variation of the action with respect to the vielbeins and the gauge field respectively,
\be\label{eq:TJvariation}
T^\mu_{\alpha}= \frac{1}{a b}\frac{\delta I_{\mathrm{EFT}}}{\delta  e^{\alpha}_\mu},\quad J^\mu= \frac{1}{a b}\frac{\delta I_{\mathrm{EFT}}}{\delta  A_\mu}.
\ee
Much of this paper will amount to an analysis of the types of terms we can write down in $\mathcal{L}$ and the consequences on $T^\mu_\alpha$ and $J^\mu$.

\subsection{Rotational symmetry}
Before we start to write down $\mathcal{L}$, however, there are two more important things to discuss.  We begin with a discussion of the consequences of discrete rotational symmetry.

For a system to preserve the rotational symmetry group $G$, the action (a functional) must be invariant under (unitary) transformations of the fields $\psi_i$.  In addition to the ``hydrodynamic" symmetries listed above, we also wish to impose some (generally discrete) rotational symmetry group $G$.  In what follows, we will assume that there is no boost-like symmetry mixing time and space; the only spacetime symmetries of interest will include time-reversal (possibly only in tandem with inversion), and the rotational group $G$.

Let $V$ denote the $d$-dimensional ``vector" representation of the group $G$ that position and momentum transform in.  We propose that the spatial vielbein indices $bc\cdots$ are the indices which will lie precisely in $V$.  Let $V(g)$ denote the unitary transformation corresponding to the group element $g\in G$, in the representation $V$.  Then the action must be invariant under the transformation (e.g.) 
\begin{equation}
    e_{s\mu}^b \rightarrow V(g)^{bc}e^c_{s\mu}.
\end{equation}
More generally, all vielbein indices transform similarly.

It is useful to build the action by classifying all possible group invariant tensors with indices $bc\cdots$.  We can then build an action by simply making sure that all vielbein indices are contracted with one of our invariants.  We will often denote these invariant tensors with the letter $f$: \begin{equation}
    f^{b_1\cdots b_n} = V(g)^{b_1c_1}\cdots V(g)^{b_nc_n}f^{c_1\cdots c_n}.
\end{equation}
Such a tensor exists if and only if $V^{\otimes n}$ contains the trivial representation.  Each copy of the trivial representation corresponds to a different invariant tensor.


Let us start by considering the example of the non-boost invariant system, but with otherwise the full rotational symmetry group $\mathrm{O}(d)$.  Here, the underlying geometry is the Aristotelian geometry \cite{Boer1,Armas2021}. The temporal vielbein $e^0_\mu$ is independent as there is no boost symmetry relating it to the spatial ones; while the spatial vielbeins are required to form a ``metric'' $h_{\mu\nu}$, which is a rank-$d$ $(d+1)\times (d+1)$ symmetric tensor.  This metric is best understood by simply writing it as \begin{equation}
    h_{\mu\nu} = e^b_\mu e^c_\nu \delta_{bc}, \label{eq:aristotleh}
\end{equation}
with $\delta_{bc}$ the \emph{unique} invariant tensor (up to taking tensor products with itself) for $\mathrm{O}(d)$.  Because all higher rank invariant tensor are simply products of Kronecker $\delta$s, there is no reason to introduce $e^b_\mu$ in the Aristotelian geometry; it suffices to simply couple systems to $h_{\mu\nu}$, the unique invariant object.   Of course, even this Aristotelian geometry itself came from breaking the Lorentz group $\mathrm{O}(1,d)$ to $\mathrm{O}(d)$.  Because the Lorentz group had unique invariant $\eta^{\alpha\beta}$, one can only couple to the spacetime metric \begin{equation}
    g_{\mu\nu} = e^\alpha_\mu e^\beta_\nu \eta_{\alpha\beta}.
\end{equation}
But with the breaking of boost symmetry, $g_{\mu\nu}$ is no longer the most general kind of background:  instead we need to keep track of the four objects $h_{\mu\nu}$, $h^{\mu\nu}$, $e^0_\mu$ and $e_0^\mu$, obeying the constraints \begin{equation}
    h_{\mu\nu}e_0^\mu = 0, \;\;\;\; h^{\mu\nu}e^0_\mu = 0, \;\;\;\; e^0_\mu e_0^\mu = -1, \;\;\;\; h_{\mu\rho}h^{\rho \nu} = \delta^\nu_\mu - e^0_\mu e_0^\nu.
\end{equation}
But, this seemingly complicated construction is greatly simplified by noticing that all of these identities follow directly from the vielbein identities in \eqnref{eq:vielbeinid}, together with (\ref{eq:aristotleh}).


It is now straightforward to deduce what happens when the rotational symmetry $\mathrm{O}(d)$ is broken further.  As explained above, there will generally be new invariant tensors to contract vielbein into.  An instructive example is a system with rectangular symmetry group $\mathrm{D}_2$, only invariant under $x\rightarrow \pm x$ and $y\rightarrow \pm y$.  In this theory, there is an invariant tensor $f$ corresponding to the Pauli $z$-matrix.  All invariants can be built out of products of $\delta$ and $f$, as can be deduced by noting that $\delta\pm f$ correspond to projections onto even numbers of $x/y$ indices.  In this $\mathrm{D}_2$-invariant theory, we can include terms proportional to each of \begin{equation}
    h_{\mu\nu} = \delta^{bc}e^b_\mu e^c_\nu, \;\;\;\; f_{\mu\nu} = f^{bc}e^b_\mu e^c_\nu
\end{equation}
in our effective action.


A particular focus of this paper will be on fluids in two spatial dimensions.  All possible point groups are classified by either $\mathbb{Z}_N$ (a discrete rotational group without parity symmetry), a dihedral group $\mathrm{D}_{N}$\footnote{Since there are totally $2N$ group elements, it is sometimes denoted as $\mrmD_{2N}$ in the math literature.} which is a semidirect product of $\mathbb{Z}_N$ with parity, and of course the two continuous groups O(2) and SO(2).  We will focus on the group $\mrmD_N$ in this paper, with a brief discussion of $\mathbb{Z}_N$ in \secref{sec:parity}.  This group consists of a rotation $r$ around a fixed point by the angle $\theta = 2\pi/N$ and a reflection $s$ around a fixed symmetry axis, i.e.
\begin{equation}
    \mrmD_N = \langle s,r|s^2 = r^N =1, s r s = r^{-1} \rangle.
\end{equation}
The group is non-abelian when $N\geq 3$, while abelian when $N\leq 2$. The construction of invariant tensors based on branching rules is well reviewed in \cite{Cook2019, Cook2021}, and we summarize the results in \tabref{tab:tensors}. 


\begin{table}[t]
\centering
\begin{tabular}{ |c|c|c|c| } 
\hline
& $\mrmD_2$ & $\mrmD_3$ & $\mrmD_4$ \\
\hline
\multirow{3}{4em}{parity-even} & $\delta_{ij}$, $\sigma^z_{ij}$ & $\delta_{ij}$, $\delta_{ix}\sigma^z_{jk}-\delta_{iy}\sigma^x_{jk}$ & $\delta_{ij}$ \\ 
& $\delta_{ij}\delta_{kl}$,  $\epsilon_{ij}\epsilon_{kl}$, $\sigma^x_{ij}\sigma^x_{kl}$, $\sigma^z_{ij}\sigma^z_{kl}$ & $\delta_{ij}\delta_{kl}$,  $\epsilon_{ij}\epsilon_{kl}$ & $\delta_{ij}\delta_{kl}$,  $\epsilon_{ij}\epsilon_{kl}$ \\ 
& $\sigma^z_{ij}\delta_{kl}\pm\delta_{ij}\sigma^z_{kl}$, $\sigma^x_{ij}\epsilon_{kl}\pm \epsilon_{ij}\sigma^x_{kl}$, & $\sigma^x_{ij}\sigma^x_{kl}+\sigma^z_{ij}\sigma^z_{kl}$  & $\sigma^x_{ij}\sigma^x_{kl}$, $\sigma^z_{ij}\sigma^z_{kl}$ \\ 
\hline
\multirow{3}{4em}{parity-odd} & $\epsilon_{ij}$, $\sigma^x_{ij}$ & $\epsilon_{ij}$ & $\epsilon_{ij}$\\ 
& $\epsilon_{ij}\delta_{kl}\pm\delta_{ij}\epsilon_{kl}$, $\sigma^x_{ij}\delta_{kl}\pm\delta_{ij}\sigma^x_{kl}$ & $\epsilon_{ij}\delta_{kl}\pm\delta_{ij}\epsilon_{kl}$ & $\epsilon_{ij}\delta_{kl}\pm\delta_{ij}\epsilon_{kl}$\\ 
& $\epsilon_{ij}\sigma^z_{kl}\pm\sigma^z_{ij}\epsilon_{kl}$, $\sigma^x_{ij}\sigma^z_{kl}\pm\sigma^z_{ij}\sigma^x_{kl}$ & $\sigma^x_{ij}\sigma^z_{kl}-\sigma^z_{ij}\sigma^x_{kl} $ & $\sigma^x_{ij}\sigma^z_{kl}\pm\sigma^z_{ij}\sigma^x_{kl} $ \\ 
\hline
\end{tabular}
\caption{ Lists of invariant $n$-tensors for various discrete rotational groups even or odd under parity symmetry. The 3-tensor for $\mrmD_3$ is chosen to respect $\mP_y$; see the main text.}
\label{tab:tensors}
\end{table}

\subsection{The classical limit and the physical spacetime}\label{sec:cl}
The other important issue to address is the classical limit of this quantum effective theory framework, corresponding to the limit $\hbar\to 0$. This is a suppression of loop corrections to quantum field theory while maintaining the classical statistical fluctuations required by the fluctuation-dissipation theorem. Schematically, we take this limit as follows \cite{Crossley2017}
\begin{equation}
    \Lambda_r \to \Lambda_r, \quad \Lambda_a \to \hbar \Lambda_a,\quad \hbar\to 0.
\end{equation}
Hence, we can write various external and dynamical fields as
\be
e^{\alpha}_{1\slash2,\mu} =e^{\alpha}_\mu\pm \frac{\hbar}{2}e^{\alpha}_{a,\mu},\quad X_{1\slash2}^\mu = X^\mu\pm\frac{\hbar}{2}X_a^\mu,\quad A_{1\slash2,\mu} = A_\mu\pm\frac{\hbar}{2}A_{a,\mu},\quad \phi_{1\slash2} = \phi \pm\frac{\hbar}{2}\phi_{a}  .
\ee
We find the gauge invariant variables as
\begin{equation}
    e^{\alpha}_{1,A} = e^{\alpha}_{A}  +\frac{\hbar}{2} e^{\alpha}_{a, A},\quad B_{1,A} = B_{A}+\frac{\hbar}{2} B_{a,A},
\end{equation}
where
\begin{align}
e^{\alpha}_{A} &= \p_A X^\mu e^{\alpha}_\mu, &  e^{\alpha}_{a,A} &= \p_A X^\mu E^{\alpha}_{a,\mu},&  E^{\alpha}_{a,\mu} &= e^{\alpha}_{a,\mu}+\mL_{X_a}e^{\alpha}_\mu, \nonumber \\
B_{A} &= \p_A X^\mu A_\mu+\p_A \phi,& B_{a,A} &= \p_A X^\mu C_{a,\mu},& C_{a,\mu} &= A_{a,\mu}+\p_\mu \phi_a+\mL_{X_a}A_\mu,
\end{align}
where $\mL_\xi$ is the Lie derivative with respect to the vector $\xi^\mu$.

The physical spacetime is defined by one copy of the vielbeins $e^{\alpha}_\mu$ and coordinates $X^\mu$. It is often more useful to then think of $\sigma^A(x)$ as the dynamical field by inverting the function $X^\mu(\sigma)$, and -- to connect with more standard notation -- just writing lower case $x^\mu$ instead of $X^\mu$.  The $a$-fields describe noise and statistical fluctuations, and are independent of $r$-fields.
Since the invariant $a$-variables are organized into $E^{\alpha}_{a,\mu }$ and $C_{a,\mu}$, we can write the Lagrangian as, to the second order,
\be\label{eq:Lcl}
\mL=  T^\mu_{\alpha} E^{\alpha}_{a,\mu }+ J^\mu C_{a,\mu}+\i  W^{\mu\nu}_{\alpha\beta}E^{\alpha}_{a,\mu}E^{\beta}_{a,\nu}+2\i  Y^{\mu\nu}_{\alpha}E^{\alpha}_{a,\mu}C_{a,\nu}+\i Z^{\mu,\nu}C_{a,\mu}C_{a,\nu} + \cdots .
\ee
The first two terms precisely correspond to the stress tensor and current in \eqnref{eq:TJvariation}. The equation of motion in the absence of stochastic fluctuations is obtained by varying $\mL_{\mathrm{cl}}$ with respect to $X_a^\mu$ and $\phi_a$ and setting $X_a^\mu = \phi_a =0$ afterwards, which means that only the leading order in $a$-fields is involved. This leads to
\begin{equation}
     e^{-1}\p_\mu\left( e T^\mu_{\alpha}\right) e^{\alpha}_\nu -  G^{\alpha}_{\nu\mu}T^\mu_{\alpha} -  F_{\nu\mu}J^\mu =0,\quad e^{-1}\p_\mu \left(e J^\mu \right)=0.
\end{equation}
We see that besides the normal Lorentz force $F_{\mu\nu}J^\nu$, there is another     Lorentz-like force $G^{\alpha}_{\mu\nu}T^\nu_{\alpha}$ induced by the torsional spacetime \cite{Bradlyn2015}.

In the classical limit and physical spacetime, taking $\Theta = \mathcal{IT}$\footnote{In $d=2$, we employ a combination of inversion and time reversal symmetry. We discuss other symmetries below.}, the KMS symmetry transformation \eqnref{eq:tildeoperator} becomes \cite{Glorioso2017} (note that $B_\mu = A_\mu + \partial_\mu \phi$; in general, we will use the same letter with multiple types of indices when the transformation between frames is standard)
\be
\tilde{E}^{\alpha}_{a,\mu}(-x) =E^{\alpha}_{a,\mu}(x) +\i \mL_{\beta^\mu}e^{\alpha}_{\mu}(x),\quad \tilde{C}_{a,\mu}(-x)  = C_{a,\mu}(x)+\i \mL_{\beta^\mu} B_\mu(x),
\ee
where \begin{equation}
    \beta^\mu(x) = \beta(x) u^{\mu}(x),
\end{equation} and the Lie derivatives read
\begin{equation}
    \mL_{\beta^\mu}e^{\alpha}_{\mu} = \p_\mu (\beta^\nu e^{\alpha}_\nu)+\beta^\nu G^{\alpha}_{\nu\mu},\quad \mL_{\beta^\mu}B_\mu = \p_\mu (\beta\mu)+\beta^\nu F_{\nu\mu}.
\end{equation}
The dynamical KMS invariance \eqnref{eq:kmscondition} therefore requires
\bes
\be\label{eq:kms0}
T^{\mu}_{\{0\}\alpha}\mL_{\beta^\mu}e^{\alpha}_{\mu}+J^\mu_{\{0\}} \mL_{\beta^\mu}B_\mu  = e^{-1}\p_\mu (e V_{\{0\}}^\mu),
\ee
\be\label{eq:kms11}
T^\mu_{\{1\}\alpha} = -W^{\mu\nu}_{\alpha\beta}\mL_{\beta^\mu} e^{\beta}_{\nu} -  Y^{\mu\nu}_{\alpha}\mL_{\beta^\mu}B_\nu,
\ee
\be\label{eq:kms12}
J^{\mu}_{\{1\}}= - Y^{\mu\nu}_{\alpha}\mL_{\beta^\mu} e^{\alpha}_{\nu}  - Z^{\mu,\nu}\mL_{\beta^\mu}B_\nu.
\ee
\ees
where the subscript $\{n\}$ denote the $n$-derivative order and $V_{\{0\}}^\mu$ is an arbitrary function with zero derivatives. The relationship with the entropy current has been discussed in \cite{GloriosoEntropy}.

\section{Thermodynamics and ideal hydrodynamics}\label{sec:ideal}
In this section we use the effective action principle to describe the thermodynamics and ideal hydrodynamics of a fluid with (discrete) rotational symmetry.  

\subsection{Factorizability}
We begin by discussing an important constraint which appears to arise from locality, KMS invariance, and the ability to couple to background gauge fields: the \emph{factorizability} of the ideal fluid action \cite{Crossley2017}.

In writing down the effective action by integrating out the UV degree of freedom as in \eqnref{eq:W2I}, we have assumed that in general it is not factorizable: namely that \begin{equation}
    I_{\mathrm{EFT}}[\Lambda] \ne I[\Lambda_1] - I[\Lambda_2].
\end{equation}  And indeed in order to describe dissipative hydrodynamics, this must be the case. However, we claim that the effective field theory must be factorizable \emph{for ideal hydrodynamics} in the models we will study in this paper:
\begin{equation}\label{eq:factorize}
    I_{\mathrm{EFT}} = I_{\mathrm{ideal}}[\Lambda_1] - I_{\mathrm{ideal}}[\Lambda_2] + \text{higher derivative terms}.
\end{equation}
KMS and locality do imply that, in (\ref{eq:W2I}), \begin{equation}
    W_{\mathrm{ideal}} = W[\Lambda_1] - W[\Lambda_2]; \label{eq:Wideal}
\end{equation} we have not found an example where (\ref{eq:Wideal}) holds while (\ref{eq:factorize}) does not.

A simple argument why factorizability is reasonable is as follows.  The lowest order in derivative action we can write down will generate thermodynamic correlators, e.g. $\langle \mathcal{O}_1(0,0) \mathcal{O}_2(0,0) \cdots \rangle$.  Now, suppose that there were a term in the ideal fluid action which could not be written in the form  (\ref{eq:factorize}): 
\begin{equation}
    \mathrm{e}^{I_{\mathrm{EFT,ideal}}} = \left\langle \mathrm{e}^{-\mathrm{i}\int  \mathcal{O}_2 \cdot \Lambda_2} \mathrm{e}^{\mathrm{i}\int  \mathcal{O}_1 \cdot \Lambda_1} \right\rangle \ne \left\langle \mathrm{e}^{-\mathrm{i}\int  \mathcal{O}_2 \cdot \Lambda_2 + \mathrm{i}\int  \mathcal{O}_1\cdot \Lambda_1} \right\rangle,
\end{equation}
where the time ordering is implicitly assumed.
Moreover, this equality would not hold even at the ideal fluid level.   This would imply that there are certain correlation functions where the ordering of operators was crucial, or in other words that there is a pair of (products of) thermodynamic operators, $\mathcal{O}$ and $\mathcal{O}^\prime$, such that \begin{equation}
    \langle [\mathcal{O},\mathcal{O}^\prime] \rangle \ne 0,
\end{equation}
i.e., the order in which we arranged these two operators in a thermal expectation value is important.  We expect that, for a conventional fluid (but with discrete rotational symmetry), such commutators can be assumed to vanish at the ideal fluid level.  As one transparent example of this, note that 
\begin{equation}
    \int \mathrm{d}^dx \langle [T^0_b(x), \mathcal{O}(y)] \rangle = \mathrm{i}\langle (\partial_b \mathcal{O})(y)\rangle = 0,
\end{equation}
as long as the thermal state is on average homogeneous.   Even more generally, it is typically the case that even if local operators fail to commute, the operators will commute in the thermodynamic limit.  A transparent example of this can be found in the hydrodynamics of a system with a non-Abelian flavor symmetry \cite{Glorioso:2020loc}:  even though the charge density operators do not commute, in the thermodynamic limit this commutator becomes very small; as a consequence, there are hydrodynamic modes for all flavor charges.


Henceforth, from here on out, we will assume (\ref{eq:factorize}) when building our ideal hydrodynamic action.  As we will see, this condition can impose non-trivial constraints on fluids with discrete rotational symmetry, which we will argue are \emph{stronger} than the constraints imposed within the conventional Landau paradigm for hydrodynamics.  As a consequence, the effective action approach  provides further constraints on the construction of a noise-free hydrodynamic theory than the conventional approach would alone.

\subsection{Thermodynamics}\label{sec:anisothermo}
We are now ready to build the ideal fluid Lagrangian.  It will generically take the form
\begin{equation}\label{eq:Lideal}
    \mL^{(1,0)} = -\varepsilon_0 b_a+p_0\chi_a+n_0 \nu_a+\pi_{0,b} \tilde{u}^{b}_a,
\end{equation}
where \begin{equation}
    \nu_a = \mu_a+b_a \mu,\quad \tilde{u}^{b}_a = u^{b}_a+b_a u^{b},
\end{equation}
and $\varepsilon_0$, $p_0$, $n_0$ and $\pi_{0,b}$ are functions of $\tau$, $\mu$ and $u^{b}$ which are not all independent (we will return to this point in Section \ref{sec:3KMS}).  It is useful to find explicit expressions for the $a$-fields in the physical spacetime, which will be of interest as we eventually use (\ref{eq:Lideal}) to deduce constitutive relations: 
\be
\beal
b_a &=u^\mu E^{0}_{a,\mu},\quad \nu_a = u^\mu C_{a,\mu},\quad \tilde{u}^{b}_a = u^\mu E^{b}_{a,\mu} ,\quad 
\chi_a=  e^\mu_{b}\left(E^{b}_{a,\mu} - u^{b} E^{0}_{a,\mu} \right),\\
V_{a,I} &= \lambda_{I}^\mu E^{0}_{a,\mu},\quad \mfc_{a,I} = \lambda_I^\mu C_{a,\mu},\quad \Xi^I_{a,J} =  a^I_{b}\lambda^\mu_J \left(E^{b}_{a,\mu} - u^{b} E^{0}_{a,\mu} \right) - \frac{1}{d}\chi_a \delta^I_J.
\eeal
\ee

There are also ``anisotropic'' terms that are allowed by \emph{all} the symmetries of Section \ref{sec:eft}:
\begin{equation}\label{eq:Laniso}
    \mL^{(1,0)}_{\mathrm{new}}=r\tau_a,
\end{equation}
where $r$ is some possibly new thermodynamic coefficient, and $\tau_a$ is some contraction of $e_{s\mu}^b$ with the $G$-invariant tensors $f$ which vanishes when $e_{a\mu}^b=0$.  The main result of this section is that \begin{equation}
    r=0. \label{eq:requals0}
\end{equation}
As such, the only anisotropy which is possible within ideal hydrodynamics arises due to the anisotropic momentum susceptibility in (\ref{eq:Lideal}).  We demonstrate this surprising fact, at least within linearized hydrodynamics, using kinetic theory models of anisotropic (electron) fluids in Appendix \ref{app:kinetic}.    Using (\ref{eq:TJvariation}), we conclude that the ideal hydrodynamic constitutive relations are \begin{subequations}
\begin{align}
    T^\mu_{0} &= -\varepsilon_0 u^\mu -p_0 (u^\mu -e^\mu_{0}), \\
    T^\mu_{b} &= p_0 e^\mu_{b}+\pi_{0,b} u^\mu, \\
    J^\mu &= n_0 u^\mu.
\end{align}
\end{subequations}

One should distinguish the anisotropy induced by the discrete rotational symmetry here from the literature where an external source has been manually added in one particular spatial direction \cite{Florkowski1, Florkowski2, Florkowski3} (see also \cite{Jain:2015txa}). In the latter case, there is a spatial Killing vector orthogonal to the temporal one $u^\mu h_\mu = 0 $, rendering the transverse and longitudinal pressure to be different.  Another example of hydrodynamics where more explicit anisotropy is possible is in the presence of a 1-form symmetry, such as in magnetohydrodynamics \cite{Grozdanov:2016tdf}.


\subsubsection{Absence of anisotropic pressure in a rectangular fluid}
It is illustrative to focus on a concrete example to justify (\ref{eq:requals0}).  Let us consider a fluid with rectangular ($\mathrm{D}_2$) point group, which will correspond to invariance under $x\rightarrow \pm x$ and $y\rightarrow \pm y$.   The invariant tensors correspond to anything with an even number of $x$ and $y$ indices, and can be built out of tensor products of $\delta_{bc}$ and $f_{bc} = (\sigma^z)_{bc}$.  In this case we will write more explicitly: 
\begin{equation}
    \mL^{(1,0)}_{\mathrm{new}} = p_{0,\times} \tau_a = p_{0,\times} f_{bc}a_{1I}^b a_{2I}^c \mathrm{``="} p_{0,\times}f^{bc}a^I_b a^J_c\Xi_{a,IJ},
\end{equation}
where the quoted equation means that the stress tensors produced by the action are equal.
Due to the tracelessness of $f_{bc}$, this term vanishes when $a_{aI}^b = 0$.  
The ideal hydrodynamic constitutive relations is then modified to
\begin{subequations}
\begin{align}
    T^\mu_{0} &= -\varepsilon_0 u^\mu -p_0 (u^\mu -e^\mu_{0}) - p_{0,\times} f^c_d e^\mu_{c}u^{d}, \\
    T^\mu_{b} &= p_0 e^\mu_{b}+p_{0,\times} f^c_b e^\mu_{c}+\pi_{0,b} u^\mu.
\end{align}
\end{subequations}
In Landau's hydrodynamic paradigm, $p_{0,\times}\ne 0$ would be possible if we could construct a conserved entropy current.  Equivalently, we can simply ask whether it is possible for (\ref{eq:kms0}) to ever hold \cite{GloriosoEntropy}. Explicitly, we have
\begin{align}
    T^\mu_{\{0\}\alpha} \mL_\beta e^\alpha_\mu = - p_{0,\times} f^c_b e^\mu_c u^b \left(\p_\mu(\beta^\nu e^0_\nu) +\beta^\nu G^0_{\nu\mu}  \right)+p_{0,\times} f^c_b e^\mu_c  \left(\p_\mu(\beta^\nu e^b_\nu) +\beta^\nu G^b_{\nu\mu}  \right).
\end{align}
In the presence of an arbitrary background vielbein, this term cannot be arranged into a total derivative.  In particular, in the presence of non-zero $G^{\alpha}_{\nu\mu}$, there are clearly terms which are not total derivatives in the above equation.  Since this term would violate KMS invariance, we must have $p_{0,\times}=0$. 

Interestingly, if we turn off the spin connection, $G^{\alpha}_{\nu\mu}=0$, and we work in the flat spacetime limit, the anisotropic pressure will be KMS invariant and consistent with the entropy current by requiring 
\begin{equation}\label{eq:thermo_0_aniso} 
    p_{0,\times} = -\beta \frac{\p p_{0,\times}}{\p \beta},
\end{equation}
where we choose $V_{\{0\}}^\mu = p_{0,\times} f^b_c e^\mu_b e^c_\nu \beta^\nu$. Therefore, to forbid the anisotropic pressure, we must impose KMS invariance upon an arbitrary backgroud field. This is analogous to the case of chiral anomaly in $1+1$d \cite{LucaPRL} where $j\sim \mu$ at leading order is fixed by KMS invariance only with generic background fields. 
Similar instances where introducing an arbitrary background field is important to fix hydrodynamic coefficients are found in \cite{Yarom2014, Boer3}.
Without appealing to KMS invariance, we show in \appref{app:integrability} that the anisotropic pressure in the action does not manifest factorizability.




\subsubsection{Absence of linear velocity in ideal stress tensor}\label{sec:factex2}
Let us now give another example of a forbidden term in a fluid with $\mathrm{D}_3$ symmetry.
This introduces a new invariant tensor which is traceless and fully symmetric: $f^{bcd}$ (see \tabref{tab:tensors}).  So it is tempting to try and write down
\begin{equation}\label{eq:Laniso2}
    \mL^{(1,0)}_{\mathrm{new}}= K_1 a^I_{2,b} f^{b}_{cd} a^{c}_{1,I} u^{d} + K_2 f^b_{cd} a^I_b u^c u^d \mathfrak{c}_{a,I}+K_3f^b_{cd} a^I_b u^c u^d V_{a,I},
\end{equation}
where 
\begin{equation}
    \mfc_{a,I}= \mfb_{a,I} +\mu V_{a,I}.
\end{equation}
By varying with respect to the background field, we find
\begin{subequations}
\begin{align}
    T^\mu_0 &= (K_3-K_1) f_{bcd} e^{b\mu}  u^c u^d, \\
    T^\mu_b &= K_1 f_{bcd} e^{c\mu} u^d, \\
    J^\mu &=K_2 f_{bcd}e^{b\mu}u^c u^d.
\end{align}
\end{subequations}
Similar to the previous example, we want to find what the KMS invariance would put as constraints. Specifically,
\begin{align}
    T^\mu_{\{0\}\alpha} \mL_\beta e^\alpha_\mu + J^\mu_{\{0\}} \mL_\beta B_\mu = & (K_3-K_1) f_{bcd} e^{b\mu}  u^c u^d \left(\p_\mu(\beta^\nu e^0_\nu) +\beta^\nu G^0_{\nu\mu}  \right)+ K_1 f_{bcd} e^{c\mu} u^d \left(\p_\mu(\beta^\nu e^b_\nu) +\beta^\nu G^b_{\nu\mu}  \right) \nonumber\\
    &+ K_2 f_{bcd}e^{b\mu}u^c u^d\left(\p_\mu(\beta\mu)+\beta^\nu F_{\nu\mu} \right).
\end{align}
Similar to above, we find that it is not possible to rewrite this term as a total derivative due to the presence of $F$  and $G$ terms; thus we must have $K_1=K_2=K_3=0$. However, when the system only couples to the flat spactime without external field, the KMS invariance condition becomes
\begin{equation}
    K_3f_{bcd}e^{b\mu}u^c u^d \p_\mu(\beta)+K_2 f_{bcd}e^{b\mu}u^c u^d \p_\mu(\beta\mu)=\frac{1}{2} f_{bcd}e^{b\mu}u^c u^d \p_\mu(\beta K_1),
\end{equation}
which leads to
\begin{equation}
    K_3+\mu K_2 = \frac{1}{2}\frac{\p(\beta K_1)}{\p\beta},\quad K_2 = \frac{1}{2} \frac{\p K_1}{\p\mu}.
\end{equation}
Indeed, without background fields, it is possible to construct a conserved entropy current for ideal hydrodynamics so long as this constraint on $K_{1,2,3}$ is satisfied.  It is only KMS invarince plus coupling to an arbitrary background field which demands $K_{1,2,3}=0$.

Should one allow such terms, the $K_1$ term would qualitatively change the dispersion relations of an ideal fluid.  In linear response about a fluid at rest, $K_{2,3}$ is a nonlinearity that will not affect quasinormal modes.

We could also try to write terms proportional to $P_2 f^b_{c} a^I_b u^c  \mathfrak{c}_{a,I}+P_3f^b_{c} a^I_b u^c  V_{a,I}$ (include $f_{bc}=\delta_{bc}$) in the action.  However, these are also not allowed.

\subsection{KMS invariance}\label{sec:3KMS}
Since very few anisotropic terms are allowed within hydrodynamics, the thermodynamic analysis will basically mirror that of a conventional fluid.  In the vielbein formalism, we must impose (\ref{eq:kms0}) to (\ref{eq:Lideal}).  We find that, with the choice $V^\mu_{\{0\}} = p_0\beta^\mu $,
\begin{equation}
    -(\varepsilon_0+p_0)\p\beta+\pi_{0,b}\p(\beta u^b)+n_0\p(\beta\mu) = \beta \p p_0,
\end{equation}
which leads to
\be\label{eq:thermo_0} 
\varepsilon_0+p_0 - \pi_{0,b} u^{b} - \mu n_0 = -\beta \frac{\p p_0}{\p \beta},\quad n_0 = \frac{\p p_0}{\p \mu},\quad   \pi_{0,b} =  \frac{\p p_0}{\p u^{b}}.
\ee
This is precisely the thermodynamic relation for a rotational invariant fluid without boost symmetry \cite{Boer1}, but with more general form of momentum susceptibilities \cite{Hartnoll_book};  e.g. in a $\mathrm{D}_2$-invariant fluid:
\begin{equation}\label{eq:momentumaniso}
    \pi_{0,b} = \rho_0 \delta_{bc} u^{c}+\rho_{0,\times} f_{bc} u^{c}+\ldots,
\end{equation}
The dots in the above equation include higher orders of velocity densities with invariant $n\geq 2$-tensors. Further, the factorizability requires the invariant tensors to be symmetric, and the combination $\pi_{0,b}u^{b}$ must be positive to ensure thermodynamic stability.

\section{First order dissipative hydrodynamics}\label{sec:dissipative}
We now turn to dissipative, first order hydrodynamics.  Unlike before, without a factorizability requirement, here we will find that nearly everything allowable by symmetry can exist.
\subsection{The effective action and transport coefficients}
The $\mO(a)$ Lagrangian with first derivatives in the fluid spacetime can be written explicitly.  We will focus on a particular example of a fluid with a symmetric traceless invariant tensor $f^{bcd}$ and tensors $f^{bcde} = f^{debc}$; however, the construction is straightforward to extend.  In particular, for most discrete groups there are multiple inequivalent such tensors, but we will postpone the full enumeration of them and their effects to later subsections (in \secref{sec:ex1} and \secref{sec:ex2}) to avoid overly cluttering the notation here. We find that
\be\label{eq:L1}
\beal
\mL^{(1,0)}+\mL^{(1,1)}=&-f_1b_a+f_2\chi_a+f_3 \nu_a+f_{4,b} \tilde{u}^{b}_a - \eta f^{IJKL} \Xi_{a,IJ}  A_{KL} \\
&- \lambda_1 V_a^I \widetilde{ D_I b_r}-\lambda_2 \mfc_a^I \hat{D}_t \mfb_{r,I} +\lambda_{12} V_a^I\hat{D}_t \mfb_{r,I} +\lambda_{21}\mfc_a^I  \widetilde{D_I b_r}\\
&+f^{IJK}\left\{ \left(\gamma_{13}  \widetilde{D_I b_r} +\gamma_{23} \hat{D}_t \mfb_{r,I}\right) \Xi_{a,JK} +\left(\gamma_{31} V_{a,I}+\gamma_{32} \mfc_{a,I}  \right) A_{JK} \right\},
\eeal
\ee
where we denoted
\begin{equation}
    f^{I_1\ldots I_n} =f^{b_1\ldots b_n} a^{I_1}_{b_1}\ldots a^{I_n}_{b_n} ,\quad A^K_L = a^K_{r,b}D_t a^{b}_{r,L},
\end{equation}
and for later convenience we introduce
\be
\begin{aligned}
    \widetilde{D_I b_r} = D_I b_r -  D_I \tau,\quad \hat{D}_t \mfb_{r,I}=D_t \mfb_{r,I} - \mu \widetilde{D_I b_r}, 
\end{aligned}
\ee
and
\bes
\begin{align}
f_1 &= \varepsilon_0+f_{11}D_t \tau +f_{12} D_t(\log a)+f_{13} \beta^{-1}D_t (\mu\beta)+f_{14,b}\beta^{-1}D_t (u^{b}\beta)+\ldots, \\
f_2 &= p_0+f_{21}D_t \tau +f_{22} D_t(\log a)+f_{23} \beta^{-1}D_t (\mu\beta)+f_{24,b}\beta^{-1}D_t (u^{b}\beta)+\ldots,\\
f_3 &= n_0+f_{31}D_t \tau +f_{32} D_t(\log a)+f_{33} \beta^{-1}D_t (\mu\beta)+f_{34,b}\beta^{-1}D_t (u^{b}\beta)+\ldots,\\
f_{4,b} &= \rho_{0,b}+f_{41,b}D_t \tau +f_{42,b} D_t(\log a)+f_{43,b} \beta^{-1}D_t (\mu\beta)+f_{44,bc}\beta^{-1}D_t (u^{c}\beta)+\ldots,
\end{align}
\ees
and $\eta$, $f$'s, and $\lambda$'s are all real functions of $\mu$, $\tau$ and $u^{b}$.
Moreover, to zeroth order in derivatives, we have
\be\label{eq:L2}
\begin{aligned}
   -\i \mL^{(2,0)}=&s_{11}b_a^2 +s_{22}\chi_a^2 +s_{33} \nu_{a}^2+2s_{12}b_a\chi_a+2s_{13}b_a\nu_a+2s_{23}\chi_a \nu_a\\
   &+2 \left(s_{14,b} b_a+s_{24,b} \chi_a+s_{34,b} \nu_a+\frac{1}{2} s_{44,bc} \tilde{u}^{c}_a \right)\tilde{u}^{b}_a\\
   &+r f^{IJKL}\Xi_{a,IJ}\Xi_{a,KL}+r_{11} V_a^I V_{a,I}+r_{22}\mfc_a^I \mfc_{a,I}+2r_{12}V_a^I\mfc_{a,I}+f^{IJK} \left(t_{13} V_{a,I}+t_{23}\mfc_{a,I}\right)\Xi_{a,JK}.
\end{aligned}
\ee

Before taking the classical limit and physical spacetime, which turns out to be more convenient, we can already see the structure of the stress tensor and current from variation of \eqnref{eq:L1}. Using \eqnref{eq:TJvariation}, we arrive at  
\begin{equation}\label{eq:stress}
\begin{aligned}
    T^\mu_{\phan\nu}\equiv & ~ T^\mu_{\alpha} e^{\alpha}_\nu\\
    =&\left(\frac{\delta \mL}{\delta b_a}- \frac{\delta \mL}{\delta u^{b}_a} u^{b} - \mu\frac{\delta \mL}{\delta \mu_a} \right)u^\mu e^{0}_\nu + \frac{\delta \mL}{\delta \chi_a}  (\delta^\mu_\nu - u^{\mu}e^{0}_{\nu}) +  \frac{\delta \mL}{\delta u^{b}_a} u^\mu e^{b}_\nu +\left(\frac{1}{b}\frac{\delta \mL}{\delta v_{a,I}}-\mu \frac{\delta \mL}{\delta \mfb_{a,I}} \right) \lambda^\mu_I e^{0}_\nu \\
    &+ \frac{\delta \mL}{\delta \Xi^I_{a,J}} \left\{\lambda^I_\rho\lambda^\mu_J(\delta^\rho_\nu - u^{\rho}e^{0}_{\nu}) - \frac{1}{d}(\delta^\mu_\nu - u^{\mu}e^{0}_{\nu})\delta^I_J\right\}.
\end{aligned}
\end{equation}
Identifying
\be\label{eq:thermovar}
\beal
&\varepsilon =- \frac{\delta \mL}{\delta b_a}+ \frac{\delta \mL}{\delta u^{b}_a} u^{b} + \mu\frac{\delta \mL}{\delta \mu_a},\quad p=\frac{\delta \mL}{\delta \chi_a},\quad \pi_{b} =  \frac{\delta \mL}{\delta u^{b}_a},\quad q^\mu = \left(\frac{1}{b}\frac{\delta \mL}{\delta v_{a,I}}-\mu \frac{\delta \mL}{\delta \mfb_{a,I}}\right)\lambda_I^\mu,
\eeal
\ee
we obtain
\begin{equation}\label{eq:stressvariation}
    T^\mu_{\phan\nu} = -\varepsilon u^\mu e^{0}_\nu+p(\delta^\mu_\nu - u^{\mu}e^{0}_{\nu})+ \pi_{b} u^\mu e^{b}_\nu +q^\mu e^{0}_\nu+\Sigma^\mu_{\phan\nu},
\end{equation}
where $\Sigma^\mu_{\phan\nu}u^\nu=0$.   
Similarly, from variation of $A_\mu$, we obtain the $U(1)$ current
\begin{equation}\label{eq:currentvariation}
    J^\mu = n u^\mu+j^\mu,
\end{equation}
where
\begin{equation}
    n = \frac{\delta \mL}{\delta \mu_{a}},\quad j^\mu  = \frac{\delta \mL}{\delta \mfb_{a,I}} \lambda_I^\mu.
\end{equation}
Explicit expressions for the above quantities are listed in \eqnref{eq:thermo} and \eqnref{eq:disscoe} in physical spacetime, as one can check they are identical to the one by variation.

Now, let us take the classical limit and work in the physical spacetime. 
The transformation of derivatives from the fluid spacetime to physical spacetime is \begin{equation}
    \partial_t = b \partial, \;\;\;\;\; \partial_I = bv_I \partial + \lambda_I^\mu \partial_\mu,
\end{equation}
where we denote \begin{equation}
    \partial = u^\mu \partial_\mu,
\end{equation}
thus
\begin{equation}
\begin{aligned}
    & D_t \tau = \p \tau,\quad  D_t (\log a) =  e^\mu_{\alpha} \p e^{\alpha}_\mu +\p_\mu u^\mu = e^{-1}\p_\mu (e u^\mu),\quad D_t (\mu\beta) = \p (\mu\beta), \\
    & \widetilde{D_I b_r} = -b \p v_I = -\frac{1}{b}\lambda^\mu_I \p_\mu b -  u^\mu \lambda^\nu_I e^{0}_{\mu\nu},\quad \hat{D}_t \mfb_{r,I} = \p (\lambda^\mu_I A_\mu)+b \mu \p v_I =\lambda^\rho_I \p_\rho u^\mu A_\mu+\lambda^\mu_I \p A_\mu+\mu b^{-1}\lambda^\rho_I \p_\rho b,\\
    & A^I_J \equiv a^I_{r,b} D_t a^{b}_{r, J} = a^I_{b}\lambda^\mu_J \left( \p e^{b}_\mu+u^\rho {\omega^{b}}_{c\rho}e^{c}_\mu+ e^{b}_\nu \overline{\p_\mu u^\nu}\right), \quad D_t (u^{b}\beta) = \p (u^{b}\beta) +u^\mu {\omega^{b}}_{c\mu} u^{c}\beta  .
\end{aligned}
\end{equation}
Then, the classical Lagrangian \eqnref{eq:Lcl} reads
\begin{subequations}
\begin{align}
    T^\mu_{0} &= -f_1 u^\mu-f_2 (u^\mu-e^\mu_{0}) +\lambda_1 w^\mu + \lambda_{12} \mfw^\mu +\eta  \tau^\mu_{b} u^{b}+p_\rho f^{\rho\nu\mu} e^{b}_\nu \delta_{bc} u^{c} + \gamma_{31} f^{\mu\nu\rho} e^{b}_\nu \delta_{bc} \left(\p e^{c}_\rho+e^{c}_\lambda \overline{\p_\rho u^\lambda}\right),\\
    T^\mu_{b} &= f_2 e^\mu_{b} + f_{4,b} u^\mu -\eta \tau^\mu_{b}-p_\rho f^{\rho\nu\mu} e^{c}_\nu \delta_{cb},\\
    J^\mu &= f_3 u^\mu -\lambda_{21} w^\mu-\lambda_{2} \mfw^\mu + \gamma_{32} f^{\mu\nu\rho}e^{b}_\nu \delta_{bc} \left(\p e^{c}_\rho+e^{c}_\lambda \overline{\p_\rho u^\lambda}\right), 
\end{align}
\end{subequations}
where
\bes\label{eq:thermo}
\begin{align}
\varepsilon\equiv f_1 &=\varepsilon_0+f_{11}\p \tau+f_{12}\theta+f_{13}T \p (\mu \beta)+ f_{14,b}T \p (u^{b} \beta) ,\\
p\equiv f_2 &=p_0 +f_{21}\p \tau+f_{22}\theta+f_{23}T \p (\mu \beta)+ f_{24,b}T \p (u^{b} \beta) ,\\
n\equiv f_3 &=n_0+f_{31}\p \tau+f_{32}\theta+f_{33}T \p (\mu \beta)+ f_{34,b}T \p (u^{b} \beta) ,\\
   \pi_b\equiv f_{4,b} &=\pi_{0,b}+f_{41,b}\p \tau+f_{42,b}\theta+f_{43,b}T \p (\mu \beta)+ f_{44,bc}T \p (u^{c} \beta) ,
\end{align}
\ees
with $\theta \equiv e^{-1}\p_\mu(e u^\mu)$, $w^\mu = h^{\mu\nu}w_\nu$, $\mfw^\mu =h^{\mu\nu}\mfw_\nu $ and
\begin{equation}
\begin{aligned}
    w_\mu &= \p_\mu \tau+u^\rho G^{0}_{\rho\mu},\quad \mfw_\mu = \p_\mu \mu+\mu \p_\mu \tau+u^\rho F_{\rho\mu},\quad p_\mu = \gamma_{13} w_\mu+\gamma_{23}\mfw_\mu,\\
    \tau^\mu_{b} &= f^{\phan c \phan c^\prime}_{b\phan b^\prime} e^\mu_{c} e^\rho_{c^\prime}\left(\p e^{b^\prime}_\rho+\overline{\p_\rho u^\nu} e^{b^\prime}_\nu\right) -\frac{1}{d}f^{\phan c \phan c^\prime}_{c\phan b^\prime}  e^\rho_{c^\prime}\left(\p e^{b^\prime}_\rho+\overline{\p_\rho u^\nu} e^{b^\prime}_\nu\right)e^\mu_{b}.   
\end{aligned}
\end{equation}
In the above equations, we defined
\be
f^{\mu\nu\rho} = f^{bcd} e^\mu_{b}e^\nu_{c}e^\rho_{d},\quad \overline{\p_\nu u^\mu }= \p_\nu u^\mu -u^\mu u^\rho G^{0}_{\rho\nu},
\ee
and the spatial spin connection vanishes properly due to contractions with invariant tensors.
Straightforwardly, we obtain \eqnref{eq:stressvariation} and \eqnref{eq:currentvariation} by identifying various thermodynamic quantities as in \eqnref{eq:thermo} and dissipative coefficients with
\begin{subequations}\label{eq:disscoe}
\begin{align}
    q^\mu &= \lambda_1 w^\mu+\lambda_{12} \mfw^\mu+\gamma_{31} f^{\mu\nu\rho} e^{b}_\nu \delta_{bc} \left(\p e^{c}_\rho+e^{c}_\lambda \overline{\p_\rho u^\lambda}\right),\\
    j^\mu &= -\lambda_{21} w^\mu-\lambda_{2} \mfw^\mu + \gamma_{32} f^{\mu\nu\rho}e^{b}_\nu \delta_{bc} \left(\p e^{c}_\rho+e^{c}_\lambda \overline{\p_\rho u^\lambda}\right),\\
    \Sigma^\mu_{\phan \nu} &= -\left(\eta \tau^\mu_{b} +p_\rho f^{\rho\nu\mu} e^{c}_\nu \delta_{cb}\right)e^{b}_\alpha (\delta^\alpha_\nu  - u^\alpha e^{0}_\nu).
\end{align}
\end{subequations}
Moving to the order $\mO(a^2)$, from \eqnref{eq:L2}, we obtain
\begin{subequations}\label{eq:2ndtensor}
\begin{align}
    W^{\mu\nu}_{00} &= s_{11}u^\mu u^\nu+s_{22}(u - e_{0})^\mu(u - e_{0})^\nu-2s_{12}u^{(\mu}(u - e_{0})^{\nu)}+r_{11}h^{\mu\nu}+r  \Pi^{\mu\nu}_{bc} u^{b} u^{c} -t_{13}f^{\mu\rho\nu} e^{b}_\rho \delta_{bc}  u^{c}, \\
    W^{\mu\nu}_{b0} &=-s_{22}  e^{(\mu}_{b}(u-e_{0})^{\nu)} +s_{12} u^{(\mu} e^{\nu)}_{b} +s_{14,b} u^\mu u^\nu- s_{24,b}(u-e_{0})^{(\mu}u^{\nu)} -r  \Pi^{\mu\nu}_{bc} u^{c}+t_{13}f^{\mu\rho\nu} e^{c}_\rho \delta_{cb},\\
    W^{\mu\nu}_{bc} &= s_{22} e^{(\mu}_{b} e^{\nu)}_{c} +s_{44,bc}u^\mu u^\nu+2s_{24,b} e^{(\mu}_{c}u^{\nu)}+r \Pi^{\mu\nu}_{bc},\\
    Y^{\mu\nu}_{0} &= s_{13} u^\mu u^\nu - s_{23}(u-e_{0})^{(\mu}u^{\nu)}+r_{12}h^{\mu\nu}-t_{23}f^{\mu\rho\nu} e^{b}_\rho \delta_{bc} u^{c},\\
    Y^{\mu\nu}_{b} &= s_{23} e^{(\mu}_{b} u^{\nu)} +s_{34,b}u^\mu u^\nu+t_{23}f^{\mu\rho\nu} e^{c}_\rho \delta_{cb},\\
    Z^{\mu\nu} &= s_{33}u^\mu u^\nu+r_{22}h^{\mu\nu},
\end{align}
\end{subequations}
where\footnote{We assume $f_{b_1b_2b_3b_4}=f_{b_3b_4b_1b_2}$, and we leave discussions about $f_{b_1b_2b_3b_4}=-f_{b_3b_4b_1b_2}$ to the end of this section.}
\begin{equation}
    \Pi^{\mu\nu}_{bc} = f^{\phan b^\prime \phan c^\prime}_{b\phan c}e^{(\mu}_{b^\prime}e^{\nu)}_{c^\prime}- \frac{2}{d}f^{\phan d \phan d^\prime}_{d\phan (c}e^{(\mu}_{b)}e^{\nu)}_{d^\prime}+\frac{1}{d^2}f^{\phan d \phan d^\prime}_{d\phan d^\prime}e^{(\mu}_{b}e^{\nu)}_{c}.
\end{equation}
Then, the dynamical KMS condition \eqnref{eq:kms11} and \eqnref{eq:kms12} tells that
\be\label{eq:kmspositive}
\beal
&s_{11} = f_{11}T,\quad s_{22} = - f_{22}T,\quad s_{33} = -f_{33}T , \quad s_{44,bc} = -f_{44,bc} T\\
& s_{12}=f_{12}T=-f_{21}T,\quad s_{13}=f_{13}T=-f_{31}T,\quad s_{23}=-f_{23}T=-f_{32}T,\\
&s_{14,b} = f_{14,b}T = -f_{41,b}T,\quad s_{24,b} = -f_{24,b}T = -f_{42,b}T,\quad s_{34,b} = -f_{34,b}T = -f_{43,b}T,\\
& t_{13}  = \gamma_{13} T = \gamma_{31} T,\quad t_{23} = \gamma_{23} T = \gamma_{32} T,\\
& r_{11} = -\lambda_1 T,\quad r_{22} = \lambda_2 T,\quad  r_{12}=-\lambda_{12}T = \lambda_{21}T,\quad r = \eta T.
\eeal
\ee
All the equations include both Onsager relations due to the symmetries $W^{\mu\nu}_{\alpha\beta} = W^{\nu\mu}_{\alpha\beta}$, $ Y^{\mu\nu}_{\alpha} = Y^{\nu\mu}_{\alpha}$, and $ Z^{\mu\nu}=Z^{\nu\mu}$ (which were enforced by the two $a$-fields in the noise terms), and the fluctuation-dissipation theorem relating noise to dissipative transport coefficients.

Defining
\begin{equation}
    \kappa\equiv -\lambda_1,\quad \sigma \equiv \lambda_2,\quad \alpha \equiv \lambda_{12} = -\lambda_{21},
\end{equation}
we identify them as the thermoelectric conductivities \cite{Hartnoll_book}. Since we work in the non-relativistic case, there are no relativistic Ward identities to relate these 3 coefficients. The bulk viscosity $\zeta$ is defined through field redefinition (see \appref{app:redef}) in \eqnref{eq:bulkvisc}. If there exists a rank-3 invariant tensor, we define 
\begin{equation}
    \gamma_\varepsilon \equiv \gamma_{13} = \gamma_{31},\quad \gamma_n \equiv \gamma_{23} = \gamma_{32}.
\end{equation}
We find that the thermoelectric conductivitivity matrix is now generalized to a 3-by-3 matrix with $\kappa,\sigma,\eta$ on the diagonal, and $\alpha,\gamma_\varepsilon,\gamma_n$ on the off-diagonal. The Onsager relation then says that the matrix is symmetric. As in the conventional transport theory, we expect both the transport coefficient matrix as well as the bulk viscosity to be positive semidefinite. This is realized by the unitarity \eqnref{eq:unitary}, which, together with \eqnref{eq:kmspositive}, implies that
\begin{equation}\label{eq:positive}
    \zeta \geq 0,\quad \eta \geq 0, \quad \sigma \geq 0,\quad \kappa\geq 0,\quad \alpha^2 \leq \kappa\sigma,\quad \gamma_\varepsilon^2\leq \kappa\eta,\quad \gamma_n^2\leq \sigma\eta.
\end{equation}

So far, we have focused on a microscopic theory that is symmetric under the combination of inversion and time reversal symmetry, i.e. $\Theta = \mathcal{I} \mT$.   It is possible to have different underlying symmetries. In particular, when $\Theta = \mathcal{T}$, all the above results hold true, except for the coefficients of terms involving invariant 3-tensors. We find that the dynamical KMS condition \eqnref{eq:kmscondition} with $\Theta = \mathcal{T}$ requires that $t_{13} = t_{23}= 0$, thus $\gamma_{13}=\gamma_{23}=\gamma_{31}=\gamma_{32}=0$.  As emphasized in \cite{triangle}, it appears that typically when the rotation  symmetry group does not include inversion, in practical applications to electron fluids one will choose $\Theta = \mathcal{IT}$.

We might also consider antisymmetric tensors: $f_{b_1b_2b_3b_4}= -f_{b_3b_4b_1b_2}$ (but parity-even).  An example with a $\mathrm{D}_2$ fluid could be $f_{yyxx} = -f_{xxyy}$.  These tensors does not generate terms in $\mL^{(2,0)}$, and according to the KMS invariance with the above $\Theta$, the terms in $\mL^{(1,1)}$ would also vanish. However, if we take the anti-unitary operator to be $\Theta = \mP_{x+y}\mT$ with
\begin{equation}
    \mP_{x+y}:x\to y,\quad y\to x,
\end{equation}
the terms in $\mL^{(1,1)}$ would not be constrained by KMS invariance and could realize further coefficients in non-dissipative hydrodynamics. This can be regarded as a generalization of non-dissipative parity-violating fluid (\secref{sec:parity}).  We will not consider this case further.


\subsection{Application to $\mathrm{O}(2)$ invariant fluid}\label{sec:ex1}

Let us first consider an $\mathrm{O}(2)$ invariant fluid in $d=2$ spatial dimensions as a sanity check of our formalism. 
Using the ``building block'' $\delta_{bc}$, we have $\pi_{0,b} = \rho_0\delta_{bc}u^{c}$, and so the thermodynamic relation \eqnref{eq:thermo_0} becomes
\begin{equation}
    \varepsilon_0+p_0 - \rho_{0} u^{2} - \mu n_0 = -\beta \frac{\p p_0}{\p \beta},\quad n_0 = 
    \frac{\p p_0}{\p \mu},\quad   \rho_{0} = 2\frac{\p p_0}{\p u^{2}},
\end{equation}
where $u^2\equiv \delta_{bc} u^{b}u^{c}$ is the ``squared velocity potential'' \cite{Boer1} and $\rho_0$ is known as the momentum susceptibility \cite{Hartnoll_book}. The nontrivial rank-4 invariant tensor is
\begin{equation}
    f_{b_1b_2b_3b_4}^{O(2)} = \delta_{b_1b_3}\delta_{b_2b_4}+\delta_{b_1b_4}\delta_{b_2b_3},
\end{equation}
while the tensor $\epsilon_{bb^\prime}\epsilon_{cc^\prime}$ leads to violation of angular momentum conservation. To see such constraint at the action level, it is most direct to employ the coset construction \cite{Landry2020} in the non-relativistic limit \cite{PencoPRD2014}.  The details of this are too technical to expand on here, but in a nutshell, the existence of angular momentum conservation in the coset construction will require that
$\Xi_{a,IJ}$ is a symmetric tensor.\footnote{A more exotic interpretation is that angular momentum conservation is a type of ``dipolar" charge conservation law.  Because of this, we must restrict the form of the momentum current to those functions which are compatible with this dipolar conservation law.  At sufficiently low order in derivatives, the only such possibility is to mandate the stress tensor be symmetric \cite{hart2021hidden,marvin}.}  Hence, we cannot couple to $\epsilon^{bc}\Xi_{a,bc}$.
Therefore, we find the shear viscosity tensor as 
\begin{equation}
    \Sigma^\mu_\nu = -\eta\left\{ \left(h^{\mu\sigma}\p h_{\sigma\rho}+h^{\mu\lambda}\overline{\p_\lambda u^\sigma}h_{\sigma\rho}+\Delta^\mu_{\phan\sigma}\overline{\p_\rho u^\sigma}\right)(\delta^\rho_\nu - u^\rho e^0_\nu)-\frac{2}{d}\left(e^\rho_{b}\p e^{b}_\rho+\Delta^\rho_{\phan\sigma}\overline{\p_\rho u^\sigma} \right) (\delta^\mu_\nu - u^\mu e^0_\nu)   \right\},
\end{equation}
where we used the ``metric'' of a rotationally invariant theory \eqnref{eq:aristotleh}, and defined
\begin{equation}\label{eq:isovis}
    \Delta^\mu_{\phan\nu} \equiv h^{\mu\rho}h_{\rho\nu} = \lambda_I^\mu \lambda^I_\nu = \delta^\mu_\nu-e^\mu_{0} e^{0}_\nu.
\end{equation}
In flat spacetime limit, we obtain (up to the treatment of bulk viscosity in Appendix \ref{app:redef})
\begin{equation}
    \Sigma_{ij}^{\mathrm{O}(2)} = -\eta\left(\p_i u_j+ \p_j u_i - \p_k u_k \delta_{ij}\right) = -\eta\left( \sigma^z_{ij}\sigma^z_{kl}+  \sigma^x_{ij}\sigma^x_{kl} \right)\p_k u_l .
\end{equation}
This agrees with the results in \cite{Boer2}. 

\subsection{Application to fluids with dihedral symmetry}\label{sec:ex2}

Next we turn to fluids with $\mathrm{D}_n$ point group.

As the first example, the $\mrmD_4$ group contains three rank-4 invariant tensors 
\begin{equation}
    f_{b_1b_2b_3b_4}^{\mrmD_4} =  \epsilon_{b_1b_2}\epsilon_{b_3b_4},\quad g_{b_1b_2b_3b_4}^{\mrmD_4} =  \sigma^z_{b_1b_2}\sigma^z_{b_3b_4},\quad h_{b_1b_2b_3b_4}^{\mrmD_4}= \sigma^x_{b_1b_2}\sigma^x_{b_3b_4},
\end{equation}
Taking $\pi_{g,b} = \rho_{g} g_{bc_1c_2c_3}^{\mrmD_4}u^{c_1}u^{c_2}u^{c_3}$ and $\pi_{h,b} = \rho_{h} h_{bc_1c_2c_3}^{\mrmD_4}u^{c_1}u^{c_2}u^{c_3}$, the thermodynamic relation becomes
\begin{equation}
    \varepsilon_0+p_0 - \rho_{0} u^{2} - \rho_{g} u^{4}_g- \rho_{h} u^{4}_h - \mu n_0 = -\beta \frac{\p p_0}{\p \beta},\quad n_0 =  \frac{\p p_0}{\p \mu},\quad   \rho_{0} = 2\frac{\p p_0}{\p u^{2}},\quad   \rho_{g} = 4\frac{\p p_0}{\p u^{4}_g},\quad   \rho_{h} = 4\frac{\p p_0}{\p u^{4}_h},
\end{equation}
where \begin{equation}
    u^4_g = g_{b_1b_1b_2b_3}^{\mrmD_4} u^{b_1}u^{b_2}u^{b_3}u^{b_4},\quad u^4_h = h_{b_1b_1b_2b_3}^{\mrmD_4} u^{b_1}u^{b_2}u^{b_3}u^{b_4}.
\end{equation} are the ``quartic velocity potentials'' and $\rho_{g,h}$ are corresponding  ``susceptibilities". Since there is no invariant rank-3 tensor, the shear viscosity becomes, in the flat spacetime limit,
\begin{equation}\label{eq:d4vis}
    \Sigma_{ij}^{\mrmD_4} =  - \left(\eta_\circ f_{ijkl}^{\mrmD_4}+\eta_1 g_{ijkl}^{\mrmD_4}+\eta_2 h_{ijkl}^{\mrmD_4}\right) \p_k u_l = -\left(\eta_\circ \epsilon_{ij}\epsilon_{kl}+ \eta_1 \sigma^z_{ij}\sigma^z_{kl}+ \eta_{2} \sigma^x_{ij}\sigma^x_{kl} \right)\p_k u_l ,
\end{equation}
where we identified $\eta_\circ$, $\eta_1$ and $\eta_2$ as the rotational, plus and cross viscosity \cite{Cook2021}. Note that when $\eta_1 = \eta_2 = \eta$ and $\eta_\circ = 0$, it reduces to the isotropic $\mathrm{O}(2)$ viscosity tensor \eqnref{eq:isovis}. The presence of the rotational viscosity $\eta_\circ$ breaks the symmetry of the dissipative contribution to the stress tensor: $\Sigma_{ij}^{\mrmD_4}\neq \Sigma_{ji}^{\mrmD_4}$. This is because $\Sigma_{ij}= \Sigma_{ji}$ is only guaranteed by \emph{continuous} $\mathrm{O}(2)$ rotational symmetry, and is unstable to any discrete rotational subgroups \cite{Cook2019}. On the other hand, the symmetric part of the viscosity tensor is relatively stable against rotational symmetry breaking (see a thorough group theory discussion in \cite{Cook2019, Cook2021}). More generally, for even $N$, the viscosity tensor of a $\mrmD_{N\geq 6}$-invariant fluid looks just like an $\mathrm{O}(2)$-invariant fluid, up to the rotational viscosity $\eta_\circ$ (which is nearly invisible in simple fluid flow experiments \cite{Cook2021,rao2021resolving}); while for $\mrmD_{N\leq 4}$, anisotropy appears in the symmetric viscosity as shown in \eqnref{eq:d4vis}. However, this is not true for odd $N$ as discussed below.

The invariant tensors for $\mrmD_3$ are 
\begin{equation}
   g_{b_1b_2b_3}^{\mrmD_3} = \delta_{b_1x}\sigma^z_{b_2b_3}-\delta_{b_1y}\sigma^x_{b_2b_3},\quad f_{b_1b_2b_3b_4}^{\mrmD_3} =  \epsilon_{b_1b_2}\epsilon_{b_3b_4},\quad h_{b_1b_2b_3b_4}^{\mrmD_3} =  \sigma^x_{b_1b_2}\sigma^x_{b_3b_4}+ \sigma^z_{b_1b_2}\sigma^z_{b_3b_4},
\end{equation}
The rank-3 tensor $g_{b_1b_2b_3}^{\mrmD_3}$ satisfies two properties: it is fully symmetric, $g_{b_1b_2b_3}^{\mrmD_3}=g_{b_2b_1b_3}^{\mrmD_3}=g_{b_3b_2b_1}^{\mrmD_3}$ and it is traceless $g_{bbc}^{\mrmD_3}=0$. We find that the thermodynamic relation becomes 
\begin{equation}
    \varepsilon_0+p_0 - \rho_{0} u^{2} - \rho_{\mrmD_3} u^{3} - \mu n_0 = -\beta \frac{\p p_0}{\p \beta},\quad n_0 = \frac{\p p_0}{\p \mu},\quad   \rho_{0} = 2\frac{\p p_0}{\p u^{2}},\quad   \rho_{\mrmD_3} = 3\frac{\p p_0}{\p u^{3}},
\end{equation}
where $u^3 = g_{b_1b_2b_3}^{\mrmD_3}u^{b_1}u^{b_2}u^{b_3}$ is the ``cubic velocity potential'' and $\rho_{\mrmD_3}$ is the corresponding momentum susceptibility. Now with the rank-3 invariant tensor, the shear viscosity tensor becomes, in the flat spacetime limit,
\begin{equation}
\begin{aligned}
    \Sigma_{ij}^{\mrmD_3} &=  - \left(\eta_\circ f_{ijkl}^{\mrmD_3}+\eta h_{ijkl}^{\mrmD_3}\right) \p_k u_l -g_{ijk}^{\mrmD_3}\left(\gamma_{13} T\p_k \beta+\gamma_{23} T \p_k \left(\beta \mu\right)\right)\\
    &= -\left(\eta_\circ \epsilon_{ij}\epsilon_{kl}+ \eta \left( \sigma^x_{ij}\sigma^x_{kl}+ \sigma^z_{ij}\sigma^z_{kl} \right) \right)\p_k u_l - \left(\delta_{ix}\sigma^z_{jk}-\delta_{iy}\sigma^x_{jk}\right)\left(\gamma_{\varepsilon}T\p_k \beta+\gamma_{n} T \p_k \left(\beta \mu\right) \right) .
\end{aligned}
\end{equation}
As we discussed above, the rotational viscosity $\eta_\circ$ indicates the breaking of $\mathrm{O}(2)$ symmetry; otherwise, the symmetric part of the viscosity tensor coincides with the $\mathrm{O}(2)$ invariant fluid.
However, there are new dissipative coefficients \cite{triangle} which couple gradients of temperature and density to the traceless symmetric velocity strain tensor:\begin{subequations}
\begin{align}
    T_{i0}^{\mrmD_3} & = \gamma_{\varepsilon}  \left(\delta_{ix}\sigma^z_{kl}-\delta_{iy}\sigma^x_{kl}\right) \p_k u_l ,\\
    J_i^{\mrmD_3} & =\gamma_{n}  \left(\delta_{ix}\sigma^z_{kl}-\delta_{iy}\sigma^x_{kl}\right) \p_k u_l,
\end{align}
\end{subequations}
and similarly, which cause stress in the presence of temperature or velocity gradients.  The coefficients are related by KMS, or equivalently, Onsager reciprocity.  From a representation theory perspective, this coupling is possible because $f_{bcd}$ converts a traceless symmetric tensor into a vector:  they are in the same irrep of $\mathrm{D}_3$.

\subsection{Normal modes}
\label{sec:normalmodes}

In this section we work out the linearized hydrodynamics of a discrete rotational fluid at rest ($u^b=0$ in equilibrium)\footnote{This condition simplifies the expressions; for example, the internal energy $\tilde{\varepsilon}=\varepsilon - \pi_{(i)}u^{(i)}\approx \varepsilon$ \cite{Boer1}. We expect a more sophisticated discussion would be necessary to generalize to a non-stationary fluid \cite{Novak2020}.}. The hydrodynamic normal modes are defined as non-vanishing solutions to the equation of motion \cite{Hartnoll_book, Kovtun2012}. Consider the linearization,
\begin{equation}
    u^\mu = (1,\delta u^i),\quad \mu = \mu_0+\delta\mu,\quad T = T_0+\delta T.
\end{equation}
To linear order in the perturbations, with the reference frame being the Landau frame (see \appref{app:redef}), the stress tensor and charge current read, combining \eqnref{eq:L1}, \eqnref{eq:Lideal} and \eqnref{eq:Lbulkvis},
\begin{equation}
\begin{aligned}
    T^0_{0} &= -\delta \varepsilon,\quad T^0_{i} = \rho_{0}\delta_{ij}  \delta u^j+\rho_{0,\times} f_{ij} \delta u^j,\quad J^0 = \delta n,\\
    T^i_{0}&= - (\varepsilon_0+p_0)\delta u^i +\lambda_1 \p^i \delta \tau+\lambda_{12} T_0 \p^i \delta(\mu\beta)+\gamma_\varepsilon f^{i\phan k}_{\phan j}\p_k \delta u^j,\\
    J^i &=  n_0\delta u^i-\lambda_{21} \p^i \delta \tau-\lambda_{2} T_0 \p^i \delta(\mu\beta)+\gamma_n f^{i\phan k}_{\phan j}\p_k \delta u^j,\\
    T^i_j &= \delta^i_{j} \left(\delta p - \zeta \p_k \delta u^k\right)-\eta\left(f_{j\phan k}^{\phan i\phan l}\p_l \delta u^k - \frac{1}{d}\delta^i_{j}f_{m\phan k}^{\phan m\phan l}\p_l \delta u^k \right)-(\gamma_\varepsilon \p_k \delta \tau+\gamma_n T_0 \p_k \delta (\mu\beta))f_{\phan\phan j}^{k i},
\end{aligned}
\end{equation}
where $f_{i\ldots}$'s are invariant tensors. Note that the external fields are turned off and we work in the flat spacetime limit.
We have changed the hydrodynamic variables from $\delta T$, $\delta \mu$ to $\delta \varepsilon$, $\delta n$, thus the variations should be expressed in terms of them: for example, $\delta p  = \left(\p_\varepsilon p \right)_{n} \delta \varepsilon+\left( \p_n p \right)_{\varepsilon} \delta n$, where the derivatives are taken in fixed value of the other variable, which is implicitly assumed below. Then, plugging it into the conservation laws
\begin{equation}
    \p_\mu T^\mu_{\alpha} = 0,\quad \p_\mu J^\mu=0,
\end{equation}
we obtain $d+2$ coupled equations (see \appref{app:NM}). In a rotationally invariant fluid, one has a shear mode with multiplicity $d-1$, which controls transverse momentum diffusion, a longitudinal diffusion mode, and two sound-like modes \cite{Hartnoll_book}. However, we find that such classification needs to be modified when the continuous rotational symmetry is broken. We consider $d=2$ in the following.

Let us first consider on a $\mrmD_2$-invariant fluid and consider the leading order contribution to the sound modes. Taking $f_{ij} = \sigma^z_{ij}$, we obtain
\begin{equation}
    \omega = \pm v_{s}(k_x,k_y)+\ldots,
\end{equation}
where the dots include dissipative corrections of order $\mO(k^2)$. The sound velocity is anisotropic:
\begin{equation}\label{eq:vsideal}
    v_s(k_x,k_y)^2 = \left[(\varepsilon_0+p_0)\p_\varepsilon p+n_0 \p_n p\right]\left(\frac{k_x^2}{\rho_0+\rho_{0,\times}}+\frac{k_y^2}{\rho_0-\rho_{0,\times}}\right),
\end{equation}
reminiscent of similar effects in anisotropic elastic solids. 
For a general dihedral group ($\mrmD_3$ or $\mrmD_4$), the sound modes have an isotropic dispersion relation at leading order:
\begin{equation}
    \omega = \pm v_{s,0}k - \i \Gamma(k_x,k_y),
\end{equation}
where $k\equiv \sqrt{k_x^2+k_y^2}$ and the speed of sound is
\begin{equation}\label{eq:vs0}
    v_{s,0}^2 = \frac{(\varepsilon_0+p_0)\p_\varepsilon p+n_0 \p_n p}{\rho_0}.
\end{equation}
Note that this formula is equivalent to the result in a rotation-invariant fluid.

The attenuation constant generically becomes anisotropic in momenta.  In this paragraph we will include transport coefficients allowed both by $\mathrm{D}_3$ and $\mathrm{D}_4$ to save space in the formulas below, but will continue to take $\rho_{0,\times}=0$:
\begin{equation}\label{eq:attenuation}
    \Gamma(k_x,k_y) = \frac{1}{2 v_{s,0}^2 \rho_0}\left[\left(\tilde{\eta}(k_x,k_y)k^{2}+\zeta k^2\right)v_{s,0}^2+\tilde{A}k^2\right]
\end{equation}
where
\begin{equation}
    \tilde{\eta}(k_x,k_y) = \frac{\eta_1\left(k_x^4+k_2^4\right)+ 2\left(2\eta_2-\eta_1\right)k_x^2 k_y^2}{k^4},
\end{equation}
and
\begin{equation}
    \tilde{A} = \left(\lambda_1 \p_\varepsilon p -\lambda_{12}\p_n p\right)\left((\varepsilon+p)\p_\varepsilon \tau+n \p_n \tau\right)+\left(\lambda_2 \p_n p+\lambda_{12}\p_\varepsilon p\right)\left((\varepsilon+p)T_0 \p_\varepsilon (\mu\beta)+n T_0 \p_n (\mu\beta) \right).
\end{equation}
We cannot detect the rotational viscosity through these modes as there is no dependence on $\eta_\circ$\footnote{Recall that we denote $\eta_\circ$, $\eta_1$ and $\eta_2$ as coefficients for invariant tensors $\epsilon_{ij}\epsilon_{kl}$, $\sigma^z_{ij}\sigma^z_{kl}$ and $\sigma^x_{ij}\sigma^x_{kl}$.}. The diffusive modes, defined as
\begin{equation}
    \omega = -\i D_{\pm} (k_x,k_y)k^2,
\end{equation}
are ``coupled together" and exhibit the rather cumbersome anisotropic diffusion constant:
\begin{equation}\label{eq:diffusion}
    D_{\pm}(k_x,k_y) = \frac{1}{2 v_{s,0}^2 \rho_0}\left(\eta(k_x,k_y)v_{s,0}^2 +A\pm \sqrt{\left(\eta(k_x,k_y)v_{s,0}^2 +A\right)^2-4v_{s,0}^2 B(k_x,k_y)}\right),
\end{equation}
where
\begin{equation}
    \eta (k_x,k_y) = \frac{(\eta_\circ+\eta_2)(k_x^4+k_y^4)+2(\eta_\circ+2\eta_1 - \eta_2)k_x^2 k_y^2}{k^4}, 
\end{equation}
\begin{equation}
    A = \left(\lambda_1 n_0 +\lambda_{12}(\varepsilon_0+p_0)\right)\left(\p_\varepsilon\tau \p_n p - \p_\varepsilon p \p_n \tau\right)+\left(\lambda_2 (\varepsilon_0+p_0)-\lambda_{12}n_0\right)\left(\p_\varepsilon p T_0\p_n (\mu\beta) - \p_n p T_0\p_\varepsilon (\mu\beta) \right),
\end{equation}
and
\begin{equation}
\begin{aligned}
    B(k_x,k_y) =& \frac{\left(3k_x^2 k_y - k_y^3\right)^2}{k^4}\Big[ \left(\gamma_n^2(\varepsilon_0+p_0)+\gamma_n\gamma_\varepsilon n\right)\left(\p_\varepsilon p T_0\p_n (\mu\beta) - \p_n p T_0\p_\varepsilon (\mu\beta) \right)\\
    &-\left(\gamma_\varepsilon^2 n_0+\gamma_n\gamma_\varepsilon(\varepsilon_0+p_0) \right)\left(\p_\varepsilon \tau \p_n p - \p_n \tau \p_\varepsilon p\right)\Big].
\end{aligned}
\end{equation}
Explicitly, the two diffusion modes are coupled through $\gamma_n$ and $\gamma_\varepsilon$ because they now carry both momentum, heat and charge. Moreover, the rotational viscosity contributes to the two diffusive modes in the same way. When $\eta_1=\eta_2=\eta$\footnote{To meet this condition, it is not necessary to have a rotationally invariant fluid, but $\mrmD_3$ or $\mrmD_{N\geq 6}$ does. Hence, it might be the reason why experiments are not able to detect the rotational viscosity even in a discrete rotational fluid.}, one can absorb the rotational viscosity into the total viscosity as $\eta_{\mathrm{eff}} = \eta_\circ+\eta$ and we recover the conventional shear mode; when $\eta_1\neq \eta_2$, the rotational viscosity is not removable. Finally, the positivity of the attenuation and diffusion constants follow from \eqnref{eq:positive} and \cite{Cook2019}. 

\section{Parity-violating hydrodynamics}\label{sec:parity}

In this section we will briefly remark on breaking the parity symmetry, which is present in $\mathrm{D}_N$.  The parity (mirror) symmetry is defined as
\begin{subequations}
\begin{align}\label{eq:parity}
    \mathcal{P}_x&: x\to -x, \quad y\to y, \\
    \mathcal{P}_y&: y\to -y, \quad x\to x.
\end{align}
\end{subequations}
Breaking the parity symmetry is fulfilled by reducing the dihedral group $\mrmD_N$ to its subgroup $\bbZ_N$ which contains only the $N$-fold rotations. 


The invariant tensors under symmetry group $\bbZ_N$ are listed in \tabref{tab:tensors} by allowing parity odd tensors to appear. Again, for the continuous symmetry group $\mathrm{SO}(2)=\mathrm{O}(2)/\bbZ_2$, angular momentum conservation will require that the action can only couple to the symmetric part of $\Xi_{a,ij}$, thus restricting the appearance of coefficients such as rotational viscosity.

In order to obtain the hydrodynamic constitutive relations, one needs to repeat the steps of the previous section, but now including these parity-violating invariant tensors. Since there are no surprises when applying this method, we will simply present the results without details of the calculation, focusing on the parity-odd coefficients which we will denote with overbars.   The equilibrium effective action is similar to \secref{sec:anisothermo} with only the momentum susceptibility becoming anisotropic by $\sigma^x_{bc}$ ($\epsilon_{bc}$ is not permitted, as it is antisymmetric). At the first derivative order, we have dissipative terms generated by symmetric tensors $\bar{f}_{bcde}=\bar{f}_{debc}$ (with $\Theta = \mT$ or $\mathcal{I T}$). However, unlike the parity-even fluids, there are non-dissipative hydrodynamics in parity-violating fluids; they are generated by anti-symmetric tensors: $\bar{f}_{bc}=-\bar{f}_{bc}$,  $\bar{f}_{bcde}=-\bar{f}_{debc}$. For the $\mathbb{Z}_N$-invariant fluid, we have:
\begin{equation}
    \bar{f}_{bc} = \epsilon_{bc}, \quad \bar{f}_{bcde} = \sigma^x_{bc}\sigma^z_{de} - \sigma^x_{de}\sigma^z_{bc},
\end{equation}
though we keep the notation more generic as the general principles would hold in other dimensions as well.  For example, we have\footnote{For simplicity we present results in the flat spacetime limit, in the stationary fluid frame, and without external electrical field.} 
\begin{equation}
   \bar{T}_{i0} = \bar{\lambda}_1 \bar{f}_{ij}\p_j \tau+\bar{\lambda}_{12} \bar{f}_{ij}\beta^{-1}\p_{j}(\mu\beta) ,\quad \bar{J}_i = \bar{\lambda}_{21} \bar{f}_{ij}\p_j \tau+\bar{\lambda}_{2} \bar{f}_{ij}\beta^{-1}\p_{j}(\mu\beta),\quad \bar{T}_{ij} = -\bar{\eta} \bar{f}_{ijkl }\p_k u_l,
\end{equation}
These antisymmetric contributions to the viscosity tensor, by construction, do not lead to any new terms in $\mL^{(2,0)}$; as such, our effective field theory will not constrain the sign or values of any parity-odd coefficients (except possibly to 0).  Note that alternative considerations \emph{can} lead to strong constraints on these coefficients \cite{tireview, Son:2009tf, Dubovsky:2011sk, Jensen:2012jy, Glorioso:2017lcn, Poovuttikul:2021mxx}.  

When $\Theta = \mT$ or $\mathcal{I T}$, we have 
\begin{equation}
    \bar{\lambda}_1 = \bar{\lambda}_2 = \bar{\lambda}_{12} = \bar{\lambda}_{21}= \bar{\eta}=0.
\end{equation}
While for $\Theta = \mP_{x,y}\mT$, all the coefficients $\bar{\lambda}_1$,$ \bar{\lambda}_2$,$\bar{\lambda}_{12}$,$ \bar{\lambda}_{21}$,$ \bar{\eta}$ are unconstrained, except for $\bar{\lambda}_{12}= \bar{\lambda}_{21}$.
In this case, $\mP_{x,y}\mT$ is the symmetry in the presence of an external magnetic field, therefore, $\bar{\lambda}_1$, $\bar{\lambda}_2$, $\bar{\lambda}_{12} = \bar{\lambda}_{21}$ correspond to the thermoelectric Hall conductivities and $\bar{\eta}$ is identified as the Hall viscosity.  Further work on anisotropic Hall viscosity can be found in \cite{BradlynPRX}.

\section{Conclusion}\label{sec:conclusion}

We have extended the effective field theory formalism proposed in \cite{Crossley2017,Glorioso2017,Liulec} to non-relativistic fluids with only discrete rotational symmetry (point group).  These fluids most naturally couple to the vielbein, whose indices transform in representations of the point group. By contracting properly the vielbein indices with invariant tensors, we are able to write down general effective actions whose variation gives the stress tensor and current; moreover, the action contains information about stochastic effects, both in and beyond linear response (though in this manuscript, we have not carefully studied the nonlinear terms in the noise $a$-fields). 

We illustrated the consequences of anisotropy on thermodynamics and first-order dissipative hydrodynamics, focusing on fluids with dihedral point groups.  We argued that due to the factorizability constraint, there will be very few new thermodynamic quantities even in highly anisotropic fluids.  In contrast, transport coefficients associated with the discrete rotational symmetry were largely unconstrained, and here our results appear to agree with earlier literature in overlapping regimes.  In tandem with a concurrent paper \cite{triangle}, we have also emphasized the possibility for novel new hydrodynamic phenomena in fluids with discrete rotational symmetries without inversion.  In these dihedral-symmetric fluids, with a small enough point group, terms that seem to be compatible with Landau's hydrodynamic phenomenology (an entropy current can be constructed) are nevertheless forbidden by factorizability, or alternatively, the ability to couple the theory to background gauge fields.  

We end with some future directions for further investigations. (\emph{1}) It is desirable to detect the anisotropic hydrodynamic phenomena we have predicted in experiments, either through viscometry \cite{Cook2021} or by careful analysis of quasinormal modes (this is likely only achievable in engineered anisotropic fluids, e.g. in liquid crystals or active matter \cite{pershan, vitelli}).  (\emph{2}) We expect generalizations of our formalism to other exotic fluids, e.g. superfluids and fracton fluids \cite{GromovPRR}, especially when including hydrodynamic fluctuations \cite{fluctuationPRL,LucaPRL}.  (\emph{3}) Studying driven open quantum systems will also be of interest\cite{SKopen}, as it is possible that factorizability is no longer a symmetry of the effective action.  (\emph{4}) It would be interesting to extend the non-relativistic anomaly \cite{JainGalileananomaly, JensenGalileananomaly} to a more general setting with the discrete rotational symmetry. We anticipate that the torsional anomaly \cite{Landsteiner2021} will induce new chiral transport coefficients in the non-relativistic limit due to the velocity density $u^b$.

\section*{Acknowledgments}

We thank Caleb Cook and Paolo Glorioso for helpful discussions and for collaboration on related work.

This work was supported in part by the Alfred P. Sloan Foundation through Grant FG-2020-13795 (AL), the National Science Foundation through CAREER Grant DMR-2145544 (AL), and through the Gordon and Betty Moore Foundation's EPiQS Initiative via Grant GBMF10279 (XH, AL).

\appendix

\section{Kinetic theory model}\label{app:kinetic}
In this section we will show that most anisotropic ideal hydrodynamic coefficients must vanish in simple kinetic theory models (without Berry curvature).  The notation and methodology in this appendix follows closely \cite{lucaskin1,lucaskin2}, and here we just provide a very terse summary.   We will study toy models of electronic Fermi liquids, in which we neglect spin.  The equilibrium distribution function is the Fermi function: \begin{equation}
    f_{\mathrm{eq}}(\mathbf{x},\mathbf{p}) = \frac{1}{1+\mathrm{e}^{\beta (\epsilon(\mathbf{p})-\mu)}},
\end{equation}
where $\epsilon(\mathbf{p})$ is the dispersion relation.  We consider a weakly perturbed system with \begin{equation}
    f = f_{\mathrm{eq}} - \frac{\partial f_{\mathrm{eq}}}{\partial \epsilon} \Phi,
\end{equation}
where $\Phi$ denotes the linearized perturbation.  If we define the inner product \begin{equation}
    \langle \Phi_1|\Phi_2\rangle = \int \frac{\mathrm{d}^d\mathbf{p}}{(2\pi \hbar)^d} \left(-\frac{\partial f_{\mathrm{eq}}}{\partial \epsilon}\right)\Phi_1\Phi_2,
\end{equation}
and the matrix \begin{equation}
    \mathsf{L} = \mathrm{i}\mathbf{k}\cdot \frac{\partial \epsilon}{\partial \mathbf{p}},
\end{equation}
then the equation of ideal hydrodynamics are simply \begin{equation}
    \partial_t |\Phi \rangle + \mathsf{L} |\Phi\rangle = 0,
\end{equation}
projected onto the hydrodynamic slow modes: charge density $|\rho\rangle$, energy density $|\epsilon\rangle$ and momentum density $|\pi_i\rangle$: \begin{subequations}
\begin{align}
    |\rho \rangle &= \int \frac{\mathrm{d}^d\mathbf{p}}{(2\pi \hbar)^d} |\mathbf{p}\rangle, \\ 
    |\epsilon \rangle &= \int \frac{\mathrm{d}^d\mathbf{p}}{(2\pi \hbar)^d} \epsilon |\mathbf{p}\rangle, \\ 
    |\pi_i \rangle &= \int \frac{\mathrm{d}^d\mathbf{p}}{(2\pi \hbar)^d} p_i |\mathbf{p}\rangle.
\end{align}
\end{subequations}
The currents within ideal hydrodynamics correspond to charge current $|J_i\rangle$, energy current $|J_i^{\mathrm{E}}\rangle$ and stress tensor $| T_{ij}\rangle$ defined as 
\begin{subequations}
\begin{align}
    |J_i \rangle &= \int \frac{\mathrm{d}^d\mathbf{p}}{(2\pi \hbar)^d} \frac{\partial \epsilon}{\partial p_i} |\mathbf{p}\rangle, \\ 
    |J_i^{\mathrm{E}} \rangle &= \int \frac{\mathrm{d}^d\mathbf{p}}{(2\pi \hbar)^d} \epsilon \frac{\partial \epsilon}{\partial p_i} |\mathbf{p}\rangle, \\ 
    |T_{ij} \rangle &= \int \frac{\mathrm{d}^d\mathbf{p}}{(2\pi \hbar)^d} p_j \frac{\partial \epsilon}{\partial p_i} |\mathbf{p}\rangle.
\end{align}
\end{subequations}

The coefficients of ideal hydrodynamics within linear response may be found as follows:  for example, \begin{equation}
    \delta J_i = \langle J_i | \rho \rangle \delta \mu + \langle J_i | \epsilon\rangle \frac{\delta T}{T} + \langle J_i | \pi_j\rangle \delta v_j,
\end{equation}
with $\delta \mu$, $\delta T$ and $\delta v_j$ corresponding to the changes in equilibrium in chemical potential, temperature and velocity respectively.  We find that \begin{subequations}
\begin{align}
    \langle J_i | \rho \rangle &= \int \frac{\mathrm{d}^d\mathbf{p}}{(2\pi \hbar)^d} \left(-\frac{\partial f_{\mathrm{eq}}}{\partial \epsilon}\right) \frac{\partial \epsilon}{\partial p_i} = 0, \\
    \langle J_i^{\mathrm{E}} | \rho \rangle = \langle J_i |\epsilon\rangle  &= \int \frac{\mathrm{d}^d\mathbf{p}}{(2\pi \hbar)^d} \left(-\frac{\partial f_{\mathrm{eq}}}{\partial \epsilon}\right) \epsilon \frac{\partial \epsilon}{\partial p_i} = 0, \\
       \langle J_i^{\mathrm{E}} | \epsilon \rangle &= \int \frac{\mathrm{d}^d\mathbf{p}}{(2\pi \hbar)^d} \left(-\frac{\partial f_{\mathrm{eq}}}{\partial \epsilon}\right) \frac{\partial \epsilon}{\partial p_i} \epsilon^2 = 0, \\
       \langle J_i | \pi_j \rangle = \langle T_{ij} |\rho\rangle  &= \int \frac{\mathrm{d}^d\mathbf{p}}{(2\pi \hbar)^d} \left(-\frac{\partial f_{\mathrm{eq}}}{\partial \epsilon}\right)  \frac{\partial \epsilon}{\partial p_i} p_j = \int \frac{\mathrm{d}^d\mathbf{p}}{(2\pi \hbar)^d} f_{\mathrm{eq}}\delta_{ij} = n_0 \delta_{ij}, \\ 
              \langle J_i^{\mathrm{E}} | \pi_j \rangle = \langle T_{ij} |\epsilon\rangle  &= \int \frac{\mathrm{d}^d\mathbf{p}}{(2\pi \hbar)^d} \left(-\frac{\partial f_{\mathrm{eq}}}{\partial \epsilon}\right)  \frac{\partial \epsilon}{\partial p_i} p_j\epsilon = \int \frac{\mathrm{d}^d\mathbf{p}}{(2\pi \hbar)^d} \left(f_{\mathrm{eq}}\epsilon \delta_{ij} + p_j \frac{\partial \epsilon}{\partial p_i}\right) = (\epsilon_0+p_0) \delta_{ij}, \\ 
                     \langle T_{ij} | \pi_k \rangle  &= \int \frac{\mathrm{d}^d\mathbf{p}}{(2\pi \hbar)^d} \left(-\frac{\partial f_{\mathrm{eq}}}{\partial \epsilon}\right)  \frac{\partial \epsilon}{\partial p_i} p_j p_k = \int \frac{\mathrm{d}^d\mathbf{p}}{(2\pi \hbar)^d} f_{\mathrm{eq}}\left(p_k\delta_{ij} + p_j \delta_{ik}\right) = \pi_{0,k} \delta_{ij}+ \pi_{0,j} \delta_{ik}.
\end{align}
\end{subequations}
where \begin{equation}
    p_0 =- \int \frac{\mathrm{d}^d\mathbf{p}}{(2\pi \hbar)^d} \log \left(1+\mathrm{e}^{-\beta \epsilon}\right).
\end{equation}
Most of the integrals above vanish because they are total derivatives of a function of $\epsilon$ integrated over a compact Brillouin zone.  Here $\pi_{0,i}$, $\rho_0$, $\epsilon_0$ and $p_0$ correspond to the background expectation values for momentum density, charge density, energy density and pressure respectively.   We conclude that within the thermodynamic currents, there are no possible contributions of the form $T_{ij} \sim f_{ijk}v_k$,  $T_{ij}\sim f_{ij}\mu$, $J_i \sim f_{ij}v_j$ etc., as noted in the main text.  

In contrast, the susceptibilities can in general be anisotropic.  For example, the inner product \begin{equation}
   \langle \pi_i | \pi_j\rangle =  \int \int \frac{\mathrm{d}^d\mathbf{p}}{(2\pi \hbar)^d} \left(-\frac{\partial f_{\mathrm{eq}}}{\partial \epsilon}\right) p_i p_j = \rho_0 \delta_{ij} + \rho_{0,\times} f_{ij} + \cdots
\end{equation}
can be as anisotropic as the dispersion relation allows.  Similarly, when some part of the vector representation is trivial, susceptibilities such as $\langle \rho|\pi_i\rangle$ can be non-vanishing.

\section{Factorizability for discrete rotational symmetry}\label{app:integrability}
In this appendix, we show that the integrability condition \begin{equation}\label{eq:integrability}
    \frac{\delta (eT^\mu_{\alpha})}{\delta e^{\beta}_\nu} =\frac{\delta (eT^\nu_{\beta})}{\delta e^{\alpha}_\mu} ,\quad \frac{\delta (eT^\mu_{\alpha})}{\delta A_\nu} =\frac{\delta (eJ^\nu)}{\delta e^{\alpha}_\mu},
\end{equation}
which is equivalent to the factorizability in \eqnref{eq:factorize} \cite{Crossley2017}, 
forbids the $p_{0,\times}$ term. To do so, it will be useful to note the following identities within ideal hydrodynamics: 
\begin{equation}
\begin{aligned}
    \delta b = b u^\mu \delta e^{0}_\mu, \quad \delta u^\mu = -u^\mu u^\nu \delta e^{0}_\nu,\quad \delta e^{\mu}_{\alpha} = -e^\nu_{\alpha}e^\mu_{\beta} \delta e^{\beta}_\nu, \quad \delta \mu = u^\mu \delta A_\mu  - \mu u^\mu \delta e^{0}_\mu.
\end{aligned}
\end{equation}
From explicit calculations, we obtain, from the second integrability condition,
\begin{subequations}
\begin{align}
    \frac{\p p_0}{\p \mu}e^\mu_{b}+\frac{\p p_{0,\times}}{\p \mu} f^{c}_{b} e^\mu_{c} &= n_0 e^\mu_{b}, \\
    \frac{\p \pi_{0,b}}{\p\mu} &= \frac{\p n_0}{\p u^{b}},\\
    \frac{\p \varepsilon_0}{\p \mu} - \mu \frac{\p n_0}{\p \mu} - u^{b} \frac{\p \pi_{0,b}}{\p \mu} &= - \frac{\p n_0}{\p \tau},
\end{align}
\end{subequations}
and we find exactly the thermodynamic relation \eqnref{eq:thermo_0} as well as the constraint $\p p_{0,\times}/\p\mu =0$.
Then the first integrability condition gives, taking $\alpha=b$ and $\beta=0$ for example,
\begin{equation}
    0 = p_{0,\times}\left(f^{c}_{c^\prime} u^{c^\prime}e_{c}^{[\nu} e_{b}^{\mu]}+f^{c}_{b} e_{c}^{[\nu} e_{0}^{\mu]} - f^{c}_{b} e_{c}^{\mu} u^\nu\right)- \frac{\p p_{0,\times}}{\p u^{b}} f^{c}_{c^\prime} e_{c}^{\mu} u^{c^\prime}u^\nu+u^{c^\prime} \frac{\p p_{0,\times}}{\p u^{c^\prime}} f^{c}_{b}u^\mu e_{c}^{\nu},
\end{equation}
where we have used the thermodynamic relation to cancel other terms. Obviously, the function $p_{0,\times}\sim T$ from the dynamical KMS condition in the flat spacetime limit does not satisfy the above equation, and the solution can only be $p_{0,\times} = 0$.

\section{Field redefinition}\label{app:redef}
In this appendix, we show how to arrive at the Landau frame by a proper field redefinition. The Landau frame is defined as
\begin{equation}
    T^\mu_{\alpha} u^{\alpha} = -\tilde{\varepsilon}_0 u^\mu,\quad J^\mu e^{0}_\mu = n_0,
\end{equation}
where $\tilde{\varepsilon}_0$ is the internal energy, and, in our case, it is modified as $\tilde{\varepsilon}_0 = \varepsilon_0- \rho_{0,b}u^{b}$.\footnote{This holds true as long as the boost symmetry is broken \cite{Boer1}.} The field redefinition consists of two parts: one by removing terms proportional to the zeroth order equation of motion, and two by proper shifting of $r$-fields
\begin{equation}
    u^\mu \to u^\mu+\delta u^\mu,\quad \beta \to \beta+\delta \beta,\quad \mu \to \mu+\delta \mu.
\end{equation}
Here we will focus on redefining the first derivative Lagrangian through (derivatives of) the zeroth order Lagrangian (for more general discussion see \cite{Glorioso2017}).  We focus on flat spacetime for simplicity.

The thermodynamic relation \eqnref{eq:thermo_0} can be written as
\begin{equation}
    \varepsilon_0 +p_0 = -\beta\left(\frac{\p p_0}{\p \beta}\right)_{\mu,u^{b}},\quad n_0 = \beta \frac{\p p_0}{\p (\mu\beta)},\quad \rho_{0,b} = \beta \frac{\p p_0}{\p (u^{b}\beta)}.
\end{equation}
Then we find the zeroth order equation of motion in flat spacetime to be
\footnote{There is also a transverse equation of motion for the stress tensor, which is not shown, as it will not be used in this discussion.}
\begin{equation}\label{eq:zeroeom}
    E_{\varepsilon} \equiv -\p \tau+\theta \left(\frac{\p p_0}{\p \varepsilon_0}\right)_{n_0,\rho_0} = 0, E_n \equiv  T \p (\mu\beta)+ \theta\left(\frac{\p p_0}{\p n_0}\right)_{\varepsilon_0,\rho_0} =0,E_{\rho}^b \equiv T\p (u^{b}\beta) +\theta\left(\frac{\p p_0}{\p \rho_{0,b}}\right)_{n_0,\varepsilon_0} = 0,
\end{equation}
where the $E_\varepsilon$, $E_n$ and $E_\rho$ are related through
\begin{subequations}
\begin{align}\label{eq:zeroeomJ}
    \p_\mu J^\mu &= -\beta\frac{\p n_0}{\p \beta} E_\varepsilon+\beta\frac{\p n_0}{\p (\mu\beta)} E_n+\beta\frac{\p n_0}{\p (u^{b}\beta)} E_\rho^b = 0,\\
    \p_\mu T^\mu_{\alpha} u^{\alpha} &= \beta\frac{\p \tilde{\varepsilon}_0}{\p \beta} E_\varepsilon-\beta\frac{\p \tilde{\varepsilon}_0}{\p (\mu\beta)} E_n-\beta\frac{\p \tilde{\varepsilon}_0}{\p (u^{b}\beta)} E_\rho^b = 0,
\end{align}
\end{subequations}
where we denoted $\ud \tilde{\varepsilon}_0 = \ud \varepsilon_0-u^{b}\ud \rho_{0,b}$ as the internal energy.
In the above equations, we used the Maxwell relations: for example, we applied
\begin{equation}
    \left(\frac{\p n_0}{\p \beta}\right)_{\mu \beta,u^{b} \beta} = -\left(\frac{\p \varepsilon_0}{\p (\mu\beta)}\right)_{\beta,u^{b}\beta},\quad  \left(\frac{\p n_0}{\p (u^{b}\beta)}\right)_{\mu \beta,\beta} = \left(\frac{\p \rho_{0,b}}{\p (\mu\beta)}\right)_{\beta,u^{b}\beta}
\end{equation}
to \eqnref{eq:zeroeomJ}. Now, let us consider the field redefinition of $r$-fields.
The leading order Lagrangian becomes
\begin{equation}
    \delta_r \mL^{(1,0)} = E_{a,\mu} \delta u^\mu+E_{a,\varepsilon_0}\delta \varepsilon_0+E_{a,\rho_{0,b}}\delta \rho_{0,b}+E_{a,n_0}\delta n_0,
\end{equation}
where $\delta \varepsilon_0 = \delta \beta \p_\beta \varepsilon_0+\delta \mu \p_\mu \varepsilon_0+\delta u^{b} \p_{u^{b}} \varepsilon_0$ (similar for $n_0$, $\rho_{0,b}$),
and with
\begin{subequations}
\begin{align}
    E_{a,\mu} &= -(\varepsilon_0+p_0) E^{0}_{a,\mu}+\rho_{0,b} E^{b}_{a,\mu}+n_0 C_{a,\mu},\\
    E_{a,\varepsilon_0} &= -\left(u^\mu+\frac{\p p_0}{\p \varepsilon_0} (u^\mu- e^\mu_{0})\right) E^{0}_{a,\mu}+\frac{\p p_0}{\p \varepsilon_0} e^\mu_{b} E^{b}_{a,\mu},\\
    E_{a,\rho_{0,b}} &= -\frac{\p p_0}{\p \rho_{0,b}} (u^\mu  - e^\mu_{0})E^{0}_{a,\mu} +    u^\mu E^{b}_{a,\mu}+\frac{\p p_0}{\p \rho_{0,b}} e^\mu_{c} E^{c}_{a,\mu},\\
    E_{a,n_0} &= u^\mu C_{a,\mu} -\frac{\p p_0}{\p n_{0}} (u^\mu  - e^\mu_{0})E^{0}_{a,\mu}+\frac{\p p_0}{\p n_0} e^\mu_{b} E^{b}_{a,\mu}.
\end{align}
\end{subequations}
Although the first equation will not be used, the coupling with the momentum susceptibility implies that one can no longer ``rotate'' $q^\mu$ into $j^\mu$ to define a frame-independent vector as in the case of relativistic fluids \cite{Crossley2017} -- this means we need a thermoelectric matrix with four transport coefficients.
Note that one part of the first derivative Lagrangian can be rewritten as
\begin{equation}\label{eq:ll}
\begin{aligned}
    \mL &= -h_\varepsilon u^\mu E^{0}_{a,\mu}+h_{\rho,b}u^\mu E^{b}_{a,\mu}+h_n u^\mu C_{a,\mu} - h_p (u^\mu-e^\mu_{0}) E^{0}_{a,\mu} + h_p e^\mu_{b} E^{b}_{a,\mu} \\
    & = -h_\varepsilon E_{a,\varepsilon_0}+h_{\rho,b}E_{a,\rho_{0,b}}+h_n E_{a,n_0}- \tilde{h}_p (u^\mu-e^\mu_{0}) E^{0}_{a,\mu} + \tilde{h}_p e^\mu_{b} E^{b}_{a,\mu},
\end{aligned}
\end{equation}
where
\begin{equation}
    \tilde{h}_p = h_p - \frac{\p p_0}{\p \varepsilon_0} h_\varepsilon-\frac{\p p_0}{\p n_0} h_n - \frac{\p p_0}{\p \rho_{0,b} } h_{\rho_{0,b}} = -\zeta \theta,
\end{equation}
and in the last step we used \eqnref{eq:zeroeom} and defined the bulk viscosity as
\begin{equation}\label{eq:bulkvisc}
\begin{aligned}
    \zeta  \equiv& f_{11} (\p_\varepsilon p)^2 - f_{22} - f_{33}(\p_n p)^2+f_{44,bc} \p_{\rho_{b}}  p \p_{\rho_{c}}  p +2 f_{12}\p_\varepsilon p -2 f_{13}\p_\varepsilon p \p_n p \\
    &-2f_{14,b}\p_\varepsilon p \p_{\rho_{b}} p +2 f_{23} \p_n p + 2 f_{24,b}\p_{\rho_{b}} p-2 f_{34,b}\p_n p\p_{\rho_{b}} p.
\end{aligned}
\end{equation}
Therefore, by eliminating the first three terms in \eqnref{eq:ll} with an appropriate choice of $\delta u^\mu$, $\delta \beta$ and $\delta \mu$, we have
\begin{equation}\label{eq:Lbulkvis}
    \mL = \zeta\theta (u^\mu-e^\mu_{0}) E^{0}_{a,\mu} - \zeta\theta e^\mu_{b} E^{b}_{a,\mu}.
\end{equation}

\section{Linearized hydrodynamics}\label{app:NM}
Here we provide a few more details for the calculations in Section \ref{sec:normalmodes}. The linearized equations of motion for hydrodynamics are
\begin{subequations}
\begin{align}
    0&=\p_0 \delta n+n_0 \p_i \delta u^i - \lambda_{21} \p^2 \delta \tau-\lambda_2 T_0 \p^2 \delta(\mu\beta)+\gamma_n f^{i\phan k}_{\phan j} \p_i \p_k \delta u^j,\\
    0&=-\p_0\delta \varepsilon - (\varepsilon_0+p_0)\p_i \delta u^i +\lambda_1 \p^2 \delta \tau+\lambda_{12} T_0 \p^2 \delta(\mu\beta)+\gamma_\varepsilon f^{i\phan k}_{\phan j} \p_i \p_k \delta u^j,\\
    0 &=\rho_0 \delta_{ij} \p_0 \delta u^j +\rho_{0,\times} g_{ij} \p_0 \delta u^j+\p_i (\delta p -\zeta \p_k \delta u^k)\nonumber \\
    &~- \eta \left(f^{\phan j \phan l}_{i\phan k\phan}\p_j \p_l \delta u^k - \frac{1}{d}f^{\phan j \phan l}_{j\phan k\phan}\p_i \p_l \delta u^k \right) -f^{kj}_{\phan\phan i}\left( \gamma_\varepsilon \p_j \p_k \delta \tau +\gamma_n T_0 \p_j \p_k \delta(\mu\beta) \right)
\end{align}
\end{subequations}
where $\p^2\equiv \p_i \p^i$. Letting
\begin{equation}
    f_{ij} = \sigma^z_{ij},\quad f_{ijk} = \delta_{ix}\sigma^z_{jk}-\delta_{iy}\sigma^x_{jk}, \quad \eta f_{ijkl} = \eta_{\circ} \epsilon_{ij}\epsilon_{kl}+\eta_1 \sigma^z_{ij}\sigma^z_{kl}+ \eta_2\sigma^x_{ij}\sigma^x_{kl},
\end{equation}
and applying Fourier transformation, we obtain a 4-by-4 matrix $M(\omega,k_x,k_y)$ acting on the vector $(\delta u^x, \delta u^y, \delta \varepsilon, \delta n)^T$ returning zero. The hydrodynamic normal modes are defined as the solutions of 
\begin{equation}
    \det M(\omega,k_x,k_y) = 0.
\end{equation}
We find
\begin{equation}
    \det M(\omega,k_x,k_y) = g_0(\mO(k^6))+ g_1(\mO(k^4))\omega+g_2(\mO(k^2))\omega^2+g_3(\mO(k^2))\omega^3+g_4(\mO(k^0))\omega^4,
\end{equation}
where $g_i(\mO(k^n))=g_i(k_x^n,k_x^{n-1}k_y,\ldots,k_y^n)$. If we substitute the ansatz $\omega = v_s (k_x,k_y)$ with $v_s (k_x,k_y)\sim \mO(k)$ into the equation, we find
\begin{equation}
    g_2(\mO(k^2))v_s^2 +g_4(\mO(k^0))v_s^4  =0.
\end{equation}
The solution is given in \eqnref{eq:vsideal}.
After finding $v_s = \pm v_{s,0}$, we take the ansatz $\omega = \pm v_{s,0} (k_x,k_y) -\i \Gamma(k_x,k_y) $ with $\Gamma(k_x,k_y)\sim \mO(k^2)$. This boils down to solving
\begin{equation}
    g_1(\mO(k^4))-\i 2\Gamma g_2(\mO(k^2)) +g_3(\mO(k^2))v_{s,0}^2  -\i 4\Gamma g_4(\mO(k^0))v_{s,0}^2  = 0.
\end{equation}
When $\rho_{0,\times}=0$, the solution is given by \eqnref{eq:attenuation} with $\rho_{0,\times}= 0$. Next, for the diffusion modes, we take the ansatz $\omega = -\i D(k_x,k_y)$ with $D(k_x,k_y)\sim \mO(k^2)$. The equation needs to be solved is
\begin{equation}
    g_0(\mO(k^6))-\i D g_1(\mO(k^4))- D^2g_2(\mO(k^2))=0,
\end{equation}
and the two solutions are given by \eqnref{eq:diffusion}.

\bibliography{PGS}

\end{document}